\documentstyle[preprint,aps,psfig]{revtex}
\tightenlines

\def\be{\begin{equation}}
\def\ee{\end{equation}}
\def\bea{\begin{eqnarray}}
\def\eea{\end{eqnarray}}

\begin{document}
\draft

\title{Relativistic Hadron-Hadron Collisions in the Ultra-Relativistic
Quantum Molecular Dynamics Model}

\author{M.~Bleicher${}^{a,e}$\thanks{E-mail:
bleicher@th.physik.uni-frankfurt.de},
E.~Zabrodin${}^{a,d}$,
C.~Spieles${}^{b,f}$,
S.A.~Bass${}^{c,f}$,
C.~Ernst${}^a$,
S.~Soff${}^{a,e}$,
L.~Bravina${}^{a,d,g}$,
M.~Belkacem${}^{a,g}$,
H.~Weber${}^a$,
H.~St\"ocker${}^a$, W.~Greiner${}^a$}

\footnotetext{${}^e$ Fellow of the Josef Buchmann Foundation}
\footnotetext{${}^f$ Feodor Lynen Fellow of the Alexander v. Humboldt
Foundation}
\footnotetext{${}^g$ Fellow of the Alexander v. Humboldt
Foundation}

\address{${}^a$ Institut f\"ur
Theoretische Physik,  J.~W.~Goethe-Universit\"at,\\
60054 Frankfurt am Main, Germany}

\address{${}^b$ Nuclear Science Division,\\
        Lawrence Berkeley National Laboratory,\\
        Berkeley, CA 94720, USA}

\address{${}^c$ Department of Physics, Duke University,\\
        Durham, N.C. 27708-0305, USA}

\address{${}^d$ Institute for Nuclear Physics,\\
Moscow State University,\\
119899 Moscow, Russia}


\maketitle

\newpage

\begin{abstract}
Hadron-hadron collisions at high
energies are investigated in the Ultra-relativistic-Quantum-Molecular-Dynamics
approach. This microscopic transport model
describes the phenomenology of hadronic
interactions at low and intermediate energies ($\sqrt s <5$~GeV) in terms
of interactions between known hadrons and their resonances. At higher
energies, $\sqrt s >5$~GeV, the excitation of color strings and their
subsequent fragmentation into hadrons dominates the multiple production of
particles in the UrQMD model.
The model shows a fair overall agreement with a large body of experimental
h-h data
over a wide range of h-h  center-of-mass energies.
Hadronic reaction data with higher precision would be useful to support the
use of the UrQMD model for relativistic heavy ion collisions. 
\end{abstract}

\newpage


\section{Motivation}
\label{sec1}

Relativistic heavy ion collision experiments at the BNL-AGS (Au(10.7~AGeV)+Au)
and at the CERN-SPS (Pb(160~AGeV)+Pb) have yielded  a large
variety of fascinating data. Various observables like the strong
$J/\Psi$ suppression, enhanced yield of intermediate mass dilepton pairs,
enhanced (anti-)hyperon yields, the creation of antimatter
clusters and strong transverse flow seem to indicate the formation of very
dense and
highly excited matter \cite{qm96}.
Since these observables are connected in a non-trivial way it is a tempting
task
for theoreticians to model high energy heavy ion
collisions in a consistent way and simultaneously predict this wide
range of observables from a few hundred MeV up to several thousand GeV per
nucleon at LHC.

Bear in mind, however, that up to now there is no unique
theoretical description of the underlying  hadron$-$hadron interactions,
with their vastly different characteristics at different incident energies
and in
different kinematic intervals. Perturbative quantum chromodynamics
(pQCD) can be applied to describe hard processes, i.e. processes
with large four-momentum, $Q^2$, transfer. But pQCD is formally
inappropriate for the description of the soft interactions because of
the absence of the large $Q^2 -$scale. Therefore,
low$-p_T$ collisions are described in terms of phenomenological
models.

Early on, multiple production of secondaries in relativistic
hadronic collisions has been described within the hydrodynamic
approach \cite{land53}. Then Regge theory
\cite{regge} and multiperipheral models have been developed  to understand
the
phenomenology of the soft
interactions. They avoid the difficulties attributed to the
statistical models. An inconvenient point of this approach is the
large number of free parameters, which have to be
fixed by comparison to experiment. Subsequently, various QCD-motivated
quark$-$parton models have been introduced.

Consequently, a vast variety of models for hadronic- and nuclear
collisions have been developed. They may be subdivided
into macroscopic (statistical and hydrodynamical) models
\cite{hydro} and microscopic (string-, transport-, cascade-, etc.)
models like, e.g. UrQMD \cite{urqmd}, which is applied in the present paper,
FRITIOF \cite{fritiof}, VENUS\cite{venus}, QGSM \cite{qgsm}, RQMD
\cite{rqmd} and others \cite{dtu,hijing,iqmd,rvuu} including the
parton cascade approach \cite{pcm}.
In the hydrodynamical (thermal) model one assumes local (global)
equilibrium - the dynamics is characterized by the equation of state employed.
The microscopic models describe subsequent individual hadron$-$hadron
collisions.

For low and intermediate energies
hadron$-$hadron and nucleus$-$nucleus collisions are described in
terms of the interactions between hadrons and their excited states,
resonances, i.e. on the (quasi-)particle level. At high energies the
quark and gluon degrees of freedom cannot be neglected. Then
the concept of color string excitations is introduced with their
subsequent fragmentation into hadrons.
In lead$-$on$-$lead collisions at the full SPS energy one finds in the
UrQMD model that the ten most frequent hadron$-$hadron collision types namely,
$N\pi, \pi\pi, \Delta\pi, NN, \pi\rho, N\Delta, \pi K, \pi\eta,
\pi\omega, \bar{K}\pi$ (in decreasing order of frequency) describe only
50\% of the total h-h collisions. The inclusion of
an additional 120 h-h collision types allows modeling up to 90\% of the
collisions predicted in the UrQMD model, while several thousand
different h-h combinations are needed to cover more than 99\% of the
total number of h-h collisions.

Since only a few of these cross sections are measured, one
relies heavily on extrapolations (and transformation, e.g. via the detailed
balance principle) of known processes.
Therefore, here we want to present and analyze the detailed elementary
h-h input used in the transport model UrQMD. The h-h predictions are a
necessary basis for our understanding of the dynamics of the  complex
heavy ion reactions.

This paper is structured as follows: a brief description of the
basic principles of the UrQMD model is given in Sec.~\ref{sec2}.
Section~\ref{sec3} presents
UrQMD-results of the different h-h cross-sections for different
reactions and a comparison with the available
experimental data. The additive
quark model (AQM) is used to calculate unknown cross-sections.
UrQMD is the first microscopic model which attempts to include
the color coherent phenomena. The implications of the
effects of color opacity and color transparency in
the model are discussed. The treatment of the formation and decay of resonances
and strings is described in detail in Sec.~\ref{sec4}. The
importance of finite size effects in the
fragmentation of strings is demonstrated. Section~\ref{sec5} discusses the
generation of the transverse momentum of particles in the model.
In Sec.~\ref{sec6} several predictions of observables for elementary
channels are presented, which are especially interesting for the upcoming
proton-proton
run of the NA49 collaboration.
Finally, a summary and  conclusions are given.

\section{The UrQMD Approach}
\label{sec2}

The UrQMD-model \cite{urqmd} is a microscopic transport theory based on the
covariant propagation of all hadrons on classical trajectories in
combination with stochastic binary scatterings, color string
formation and resonance decay. It represents a Monte Carlo solution of a
large set of coupled partial integro-differential equations for the time
evolution of the various phase space densities $f_i(x,p)$ of
particle species $i=N,\Delta,\Lambda,$ etc., which non-relativistically
assumes the Boltzmann form:
\be
\frac{{\rm d}f_i(x,p)}{{\rm d}t}\equiv
\frac{\partial p}{\partial t}\frac{\partial f_i(x,p)}{\partial p}+
\frac{\partial x}{\partial t}\frac{\partial f_i(x,p)}{\partial x}+
\frac{\partial f_i(x,p)}{\partial t}=
{\rm St} f_i(x,p)\quad,
\ee
where $x$ and $p$ are the position and momentum of the particle,
respectively, and ${\rm St} f_i(x,p)$ denotes the collision (or
rather source-) term of these
particle species, which are connected to any other particle species
$f_k$.

The exchange of electric and baryonic charge, strangeness and four momentum
in the
$t$-channel is considered for baryon-baryon (BB) collisions at low
energies, while meson-baryon (MB) and meson-meson (MM)
interactions are treated via the formation and decay of resonances, i.e. the
$s$-channel reactions. $t$-channel reactions for MB and MM collisions are
taken into
account from $\sqrt s> 3$~GeV on increasing to the only MB, MM interaction type
above $\sqrt s = 6$~GeV. For nucleus-nucleus
collisions the soft binary and ternary interactions between nucleons can be
described by the real part of the in-medium G-Matrix, which is
approximated by a non-relativistic density-dependent Skyrme potential of
the form
\be
V^{Sk} = \frac{1}{2!} t_1 \sum_{i \neq j} \delta(\vec x_i - \vec x_j)
+ \frac{1}{3!} t_2 \sum_{i \neq j \neq k} \delta (\vec x_i - \vec x_j)
\delta (\vec x_j - \vec x_k)\quad,
\ee
where $\vec x_{\alpha}$ denotes the coordinate variable in the quantum
phase space. The first term simulates the attractive potential of
the NN-interaction, and the second one yields  the saturation. In addition,
Yukawa and Coulomb potentials
are implemented in the model. The potentials allow to calculate the
equation of state of the interacting many body system, as long as it is
dominated by nucleons.
Note that these potential interactions
are only used in the model for baryons/nucleons with relative momenta
$\Delta p$ of less than 2$\,$GeV/$c$. For the hadronic collisions
discussed here, the potential interactions are omitted.
Further details of the application of the UrQMD model to heavy-ion
reactions may be found in \cite{urqmd}.

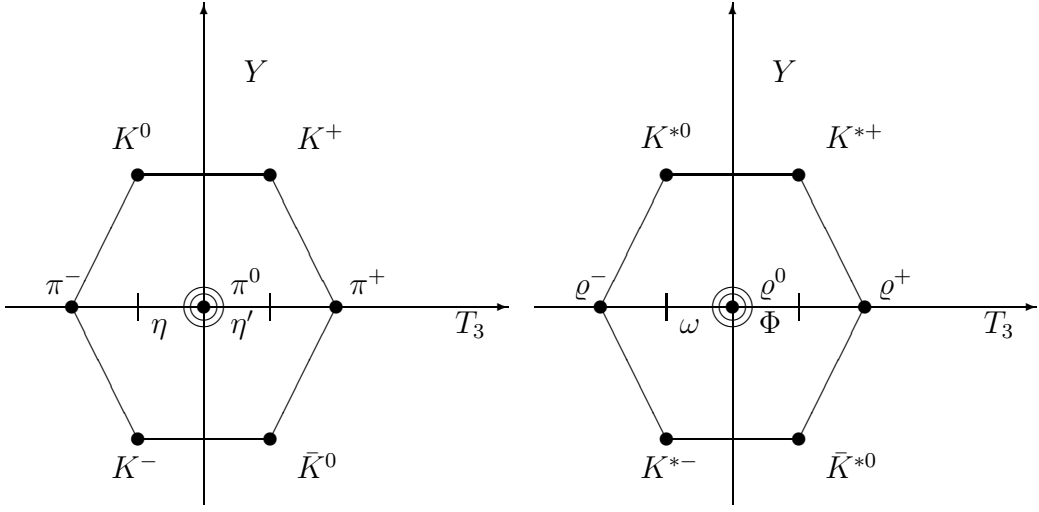
\begin{figure}
\setlength{\unitlength}{5pt}
  \begin{picture}(75,40)(0,0)
    \put(3,16){$\pi^-$}
    \put(17,16){$\pi^0$}
    \put(26,16){$\pi^+$}
    \put(17,13){$\eta '$}
    \put(11,13){$\eta$}
    \put(8,27){$K^0$}
    \put(22,27){$K^+$}
    \put(8,2){$K^-$}
    \put(22,2){$\bar K^0$}
    \put(0,15){\vector(1,0){38}}
    \put(15,0){\vector(0,1){38}}
    \put(5,15){\line(1,2){5}}
    \put(10,25){\line(1,0){10}}
    \put(20,25){\line(1,-2){5}}
    \put(25,15){\line(-1,-2){5}}
    \put(20,5){\line(-1,0){10}}
    \put(10,5){\line(-1,2){5}}
    \put(10,16){\line(0,-1){2}}
    \put(20,16){\line(0,-1){2}}
    \put(15,15){\circle*{1}}
    \put(15,15){\circle{2}}
    \put(15,15){\circle{3}}
    \put(5,15){\circle*{1}}
    \put(10,25){\circle*{1}}
    \put(20,25){\circle*{1}}
    \put(25,15){\circle*{1}}
    \put(20,5){\circle*{1}}
    \put(10,5){\circle*{1}}
    \put(34,13){$T_3$}
    \put(18,32){$Y$}
    \put(43,16){$\varrho^-$}
    \put(57,16){$\varrho^0$}
    \put(66,16){$\varrho^+$}
    \put(57,13){$\Phi$}
    \put(51,13){$\omega$}
    \put(48,27){$K^{*0}$}
    \put(62,27){$K^{*+}$}
    \put(48,2){$K^{*-}$}
    \put(62,2){$\bar K^{*0}$}
    \put(40,15){\vector(1,0){38}}
    \put(55,0){\vector(0,1){38}}
    \put(45,15){\line(1,2){5}}
    \put(50,25){\line(1,0){10}}
    \put(60,25){\line(1,-2){5}}
    \put(65,15){\line(-1,-2){5}}
    \put(60,5){\line(-1,0){10}}
    \put(50,5){\line(-1,2){5}}
    \put(50,16){\line(0,-1){2}}
    \put(60,16){\line(0,-1){2}}
    \put(55,15){\circle*{1}}
    \put(55,15){\circle{2}}
    \put(55,15){\circle{3}}
    \put(45,15){\circle*{1}}
    \put(50,25){\circle*{1}}
    \put(60,25){\circle*{1}}
    \put(65,15){\circle*{1}}
    \put(60,5){\circle*{1}}
    \put(50,5){\circle*{1}}
    \put(74,13){$T_3$}
    \put(58,32){$Y$}
  \end{picture}
\caption{
Implemented mesons: pseudo-scalar mesons (left plot) and vector
mesons (right plot). }
\label{fig1}
\end{figure}

\begin{figure}
\setlength{\unitlength}{5pt}
\begin{picture}(75,40)(0,0)
    \put(3,16){$a_0^-$}
    \put(17,16){$a_0^0$}
    \put(26,16){$a_0^+$}
    \put(17,13){$f_0$}
    \put(11,13){$f_0^*$}
    \put(8,27){$K_0^{*0}$}
    \put(22,27){$K_0^{*+}$}
    \put(8,2){$K_0^{*-}$}
    \put(22,2){$\bar K_0^{*0}$}
    \put(0,15){\vector(1,0){38}}
    \put(15,0){\vector(0,1){38}}
    \put(5,15){\line(1,2){5}}
    \put(10,25){\line(1,0){10}}
    \put(20,25){\line(1,-2){5}}
    \put(25,15){\line(-1,-2){5}}
    \put(20,5){\line(-1,0){10}}
    \put(10,5){\line(-1,2){5}}
    \put(10,16){\line(0,-1){2}}
    \put(20,16){\line(0,-1){2}}
    \put(15,15){\circle*{1}}
    \put(15,15){\circle{2}}
    \put(15,15){\circle{3}}
    \put(5,15){\circle*{1}}
    \put(10,25){\circle*{1}}
    \put(20,25){\circle*{1}}
    \put(25,15){\circle*{1}}
    \put(20,5){\circle*{1}}
    \put(10,5){\circle*{1}}
    \put(34,13){$T_3$}
    \put(18,32){$Y$}
    \put(43,16){$a_1^-$}
    \put(57,16){$a_1^0$}
    \put(66,16){$a_1^+$}
    \put(57,13){$f_1'$}
    \put(51,13){$f_1$}
    \put(48,27){$K_1^{*0}$}
    \put(62,27){$K_1^{*+}$}
    \put(48,2){$K_1^{*-}$}
    \put(62,2){$\bar K_1^{*0}$}
    \put(40,15){\vector(1,0){38}}
    \put(55,0){\vector(0,1){38}}
    \put(45,15){\line(1,2){5}}
    \put(50,25){\line(1,0){10}}
    \put(60,25){\line(1,-2){5}}
    \put(65,15){\line(-1,-2){5}}
    \put(60,5){\line(-1,0){10}}
    \put(50,5){\line(-1,2){5}}
    \put(50,16){\line(0,-1){2}}
    \put(60,16){\line(0,-1){2}}
    \put(55,15){\circle*{1}}
    \put(55,15){\circle{2}}
    \put(55,15){\circle{3}}
    \put(45,15){\circle*{1}}
    \put(50,25){\circle*{1}}
    \put(60,25){\circle*{1}}
    \put(65,15){\circle*{1}}
    \put(60,5){\circle*{1}}
    \put(50,5){\circle*{1}}
    \put(74,13){$T_3$}
    \put(58,32){$Y$}
  \end{picture}
\caption{Implemented mesons: scalar mesons (left plot) and
pseudo-vector mesons (right plot). $f_1$ and $f^\prime_1$ are the
states $f_1(1285)$ and $f_1(1420)$, respectively.
}
\label{fig2}
\end{figure}
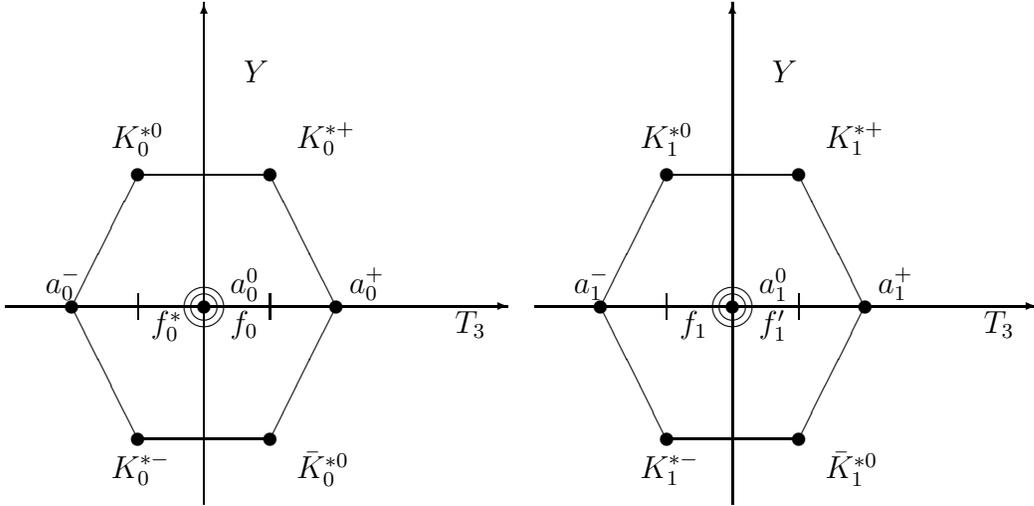

This framework allows to bridge with one concise model the entire available
range of energies from the SIS energy region ($\sqrt{s} \approx 2\,$GeV) to
the RHIC energy ($\sqrt{s} = 200\,$GeV). At the highest energies, a huge
number of
different particle species can be produced. The model should allow for
subsequent rescatterings. The collision term in the UrQMD model includes
more than fifty
baryon species and five meson nonetts (45 mesons). The
baryons and baryon resonances included in the UrQMD are listed in
Table \ref{tab1}. In addition, their antiparticles have
been implemented using charge-conjugation to assure full
baryon-antibaryon symmetry.
Figures \ref{fig1} and \ref{fig2} depict the implemented
meson multiplets: pseudo-scalar, vector, scalar, pseudo-vector and
(not shown in the Figs.) the tensor mesons as well as the heavy vector meson
resonances $\rho(1450)$, $\rho(1700)$, $\omega(1420)$, and
$\omega(1600)$. Extremely heavy meson resonances ($m>2$~GeV) are not
explicitly implemented, however they may be important when
investigating, e.g. the dynamics of $\Phi\Phi$ correlations in future
experiments.

All particles can be produced in hadron-hadron collisions and
can interact further with each other. The different decay
channels all nucleon-, $\Delta$- and hyperon-resonances up to
2.25$\,$GeV/$c^2$ mass as well as the meson (e.g. K$^*$) decays etc. are
implemented.  At higher energies we take advantage of the
hadron universality and use a string model for the decay of
intermediate states. The cross-sections of various hadronic processes
as well as the formation and fragmentation of the strings are
discussed in the subsequent chapters.

\section{Cross-Sections}
\label{sec3}

A basic input into the microscopic transport models are the particle
species and -energy dependent cross-sections of hadron-hadron interactions.
The total cross-sections are  interpreted geometrically. A collision between
two hadrons will
occur if $d<\sqrt{\sigma_{\rm tot}/\pi}$, where $d$ and $\sigma_{\rm tot}$
are the impact parameter of the hadrons and the total cross-section of the
two hadrons, respectively.
In the UrQMD model the total cross-section $\sigma_{\rm tot}$ depends on the
isospins of colliding particles, their flavor and the c.m. energy.
However, partial cross-sections are then used to calculate the relative
weights
for the different channels. Only a small fraction of all possible
hadronic cross-sections has been measured. In the following sections, we
compare the UrQMD cross-sections with experimental data. If no data
are available, the additive quark model and detailed balance
arguments are used to extrapolate such  unknown observables.

\subsection{Baryon-Baryon Cross-Sections}
\label{subsec3a}
The total BB cross-section of the reaction $A + C \rightarrow D + E$
has the general form
\be
\sigma_{tot}^{BB}(\sqrt{s}) \propto (2S_D + 1) (2S_E + 1) \frac{
\langle p_{D,E} \rangle }{\langle p_{A,C} \rangle }\, 
\frac{1}{s}\, |{\cal M}|^2 \quad,
\ee
with the spins of the particles, $S_i$, momenta of the pairs of
particles, $<p_{i,j}>$, in the two-particle rest frame, and the matrix
element $|{\cal M}|^2$. The matrix element $|{\cal M}|^2$, however,
can take on a very complicated form and may be in general a function of
of all the particle's quantum numbers as well as it's momenta and
the c.m. energy. 

If high quality experimental data on the respective cross section exists, a 
phenomenological fit to the respective data is by far the most accurate
approach for implementing the cross section. Otherwise, we have to rely
on simplified assumption for the matrix element and employ general
symmetries, like the principle of detailed balance 
(see section~\ref{res_sec}). 

\begin{figure}
\centerline{\psfig{figure=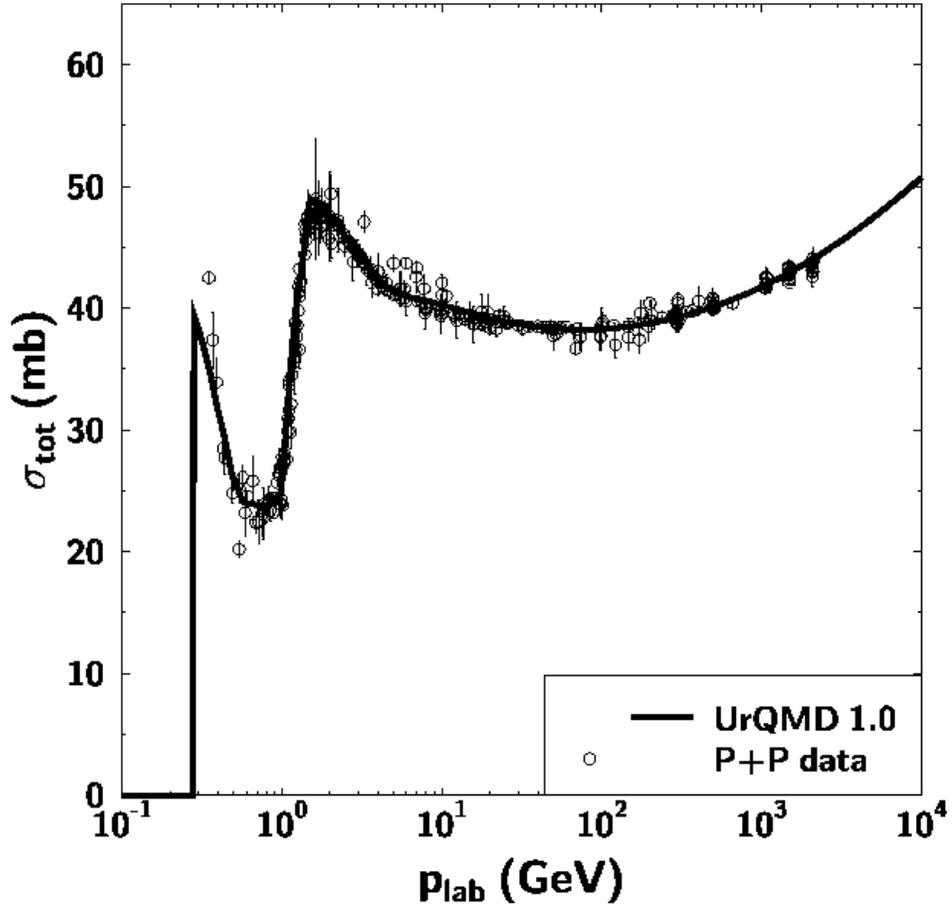,width=15cm}}
\caption{The total cross-section of $pp$ collisions  vs.
the laboratory momentum $p_{lab}$ of the incident particle.
Data are taken from \protect\cite{pdg96}.
}
\label{fig3}
\end{figure}

Let us start by investigating the total cross-section of proton-proton
collisions from a beam momentum of 0.1$\,$GeV/$c$ up to
$10^4\,$GeV/$c$ as shown in Fig.~\ref{fig3}. The total and inelastic
cross-sections of the $pp$ reaction are well measured in this energy
region \cite{pdg96}. One finds a complex structure in this
cross-section: local minima at 700$\,$MeV/$c$ ($E_{cm}\approx 2\,$GeV)
and 100$\,$GeV/$c$ ($E_{cm} \approx 10\,$GeV), the maximum at
2$\,$GeV/$c$ ($E_{cm} \approx 2-3\,$GeV) and a rise above
100$\,$GeV/$c$. Note that the steep rise in the data below
300$\,$MeV/$c$ is due to soft Coulomb interaction of the protons and is
taken care of via the potential interaction.

\begin{figure}
\centerline{\psfig{figure=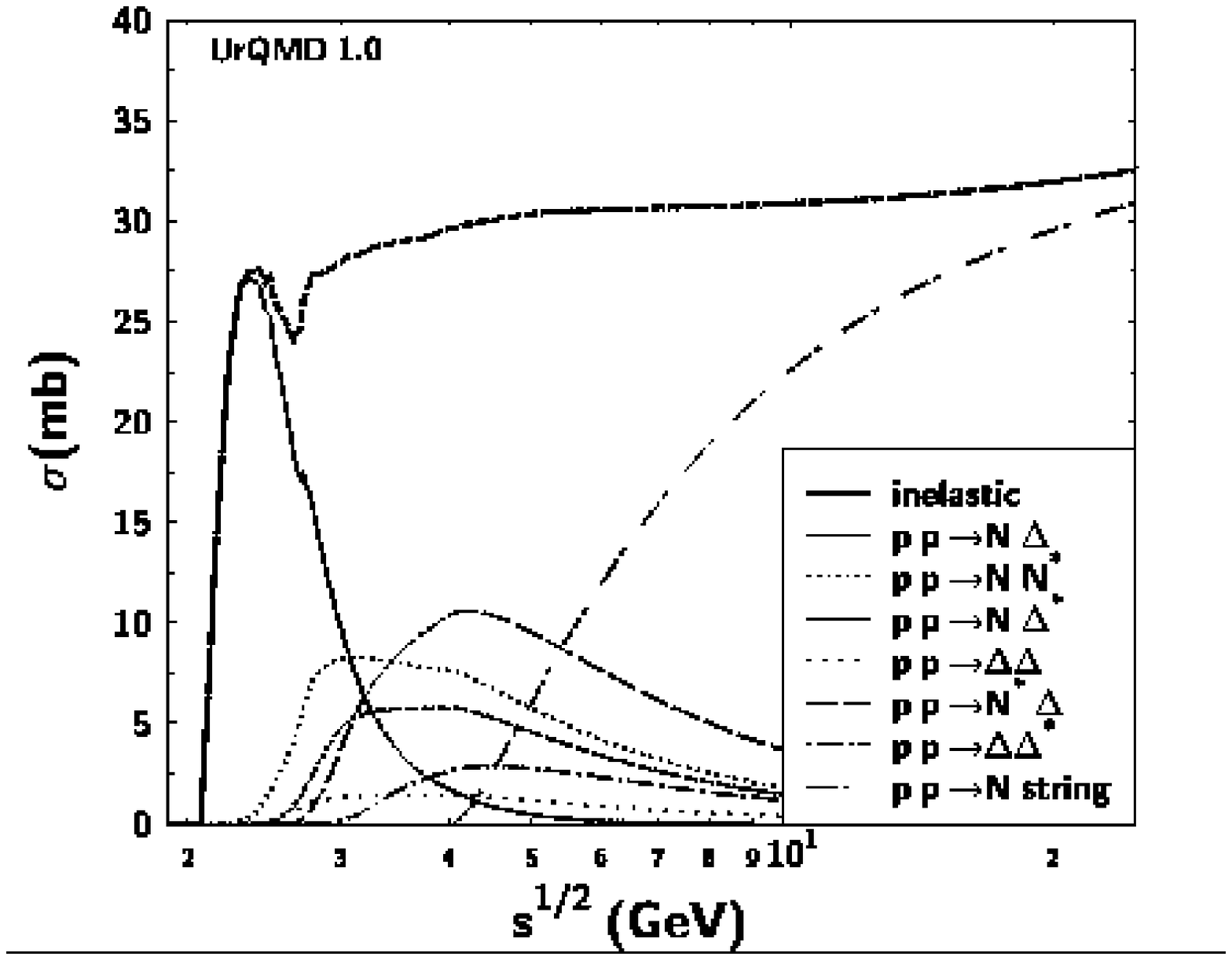,width=15cm}}
\caption{The inelastic cross-section of $pp$ collisions vs. the
laboratory momentum $p_{lab}$ and the cross-sections of the
various inelastic channels.}
\label{fig4}
\end{figure}

The structure in the $pp$ cross-section is mainly due to the inelastic
channels which are shown in Fig.~\ref{fig4}. One clearly sees
the $\Delta$ excitation with its  increasing cross-section
at low energies. 
The different partial
cross sections depicted in Fig.~\ref{fig4} are discussed
in section~\ref{res_sec}.

\begin{figure}
\centerline{\psfig{figure=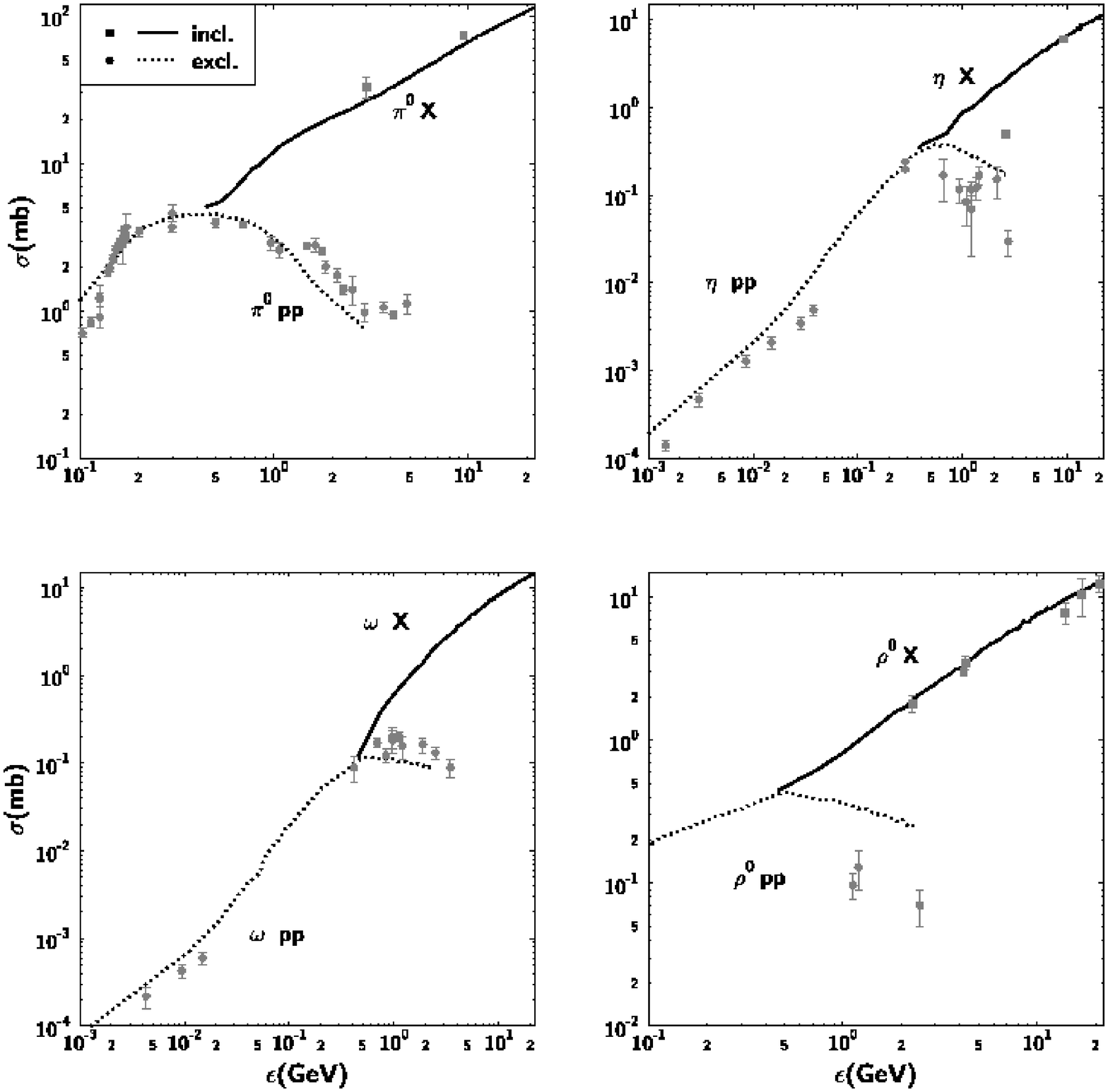,width=15cm}}
\caption{Cross section for the production of neutral mesons in $pp$. The
inclusive and exclusive meson production is compared to data by
\protect\cite{calen96a} \protect\cite{flaminio}
}
\label{ppc}
\end{figure}

A detailed comparison of prominent outgoing channels is depicted in
Fig. \ref{ppc}. Here we show calculations of
exclusive ($pp\to m pp$) and inclusive ($pp\to
mX$) cross sections for the production of
neutral mesons $m=\pi^0,\eta, \rho^0, \omega$ as a function
of the excess energies $\epsilon=\sqrt{s}-\sqrt{s_{\rm th}}$.
Here $\sqrt{s_{\rm th}}$ is the energy of the production threshold
calculated as $\sqrt{s_{\rm th}}=2m_p+m_{m}$ with the proton
mass $m_p$ and the pole mass of the meson $m_{m}$.
In the case of $\rho^0$ mesons we count only those with
masses within $\pm 100$~MeV around the pole mass to compare
with data. The exclusive $\eta$ production just
above threshold \cite{calen96a}
is overestimated by a factor of about two.
Note that above 3.5~GeV the exclusive cross sections become
less important because the string and multiple decay channels
open and allow for multiple resonance production.
For upcoming GSI-SIS experiments relevant $\sqrt{s}$ values are below
4~GeV, where so far no data on the inclusive channels
are available and therefore rely heavily on extrapolations.

\begin{figure}
\centerline{\psfig{figure=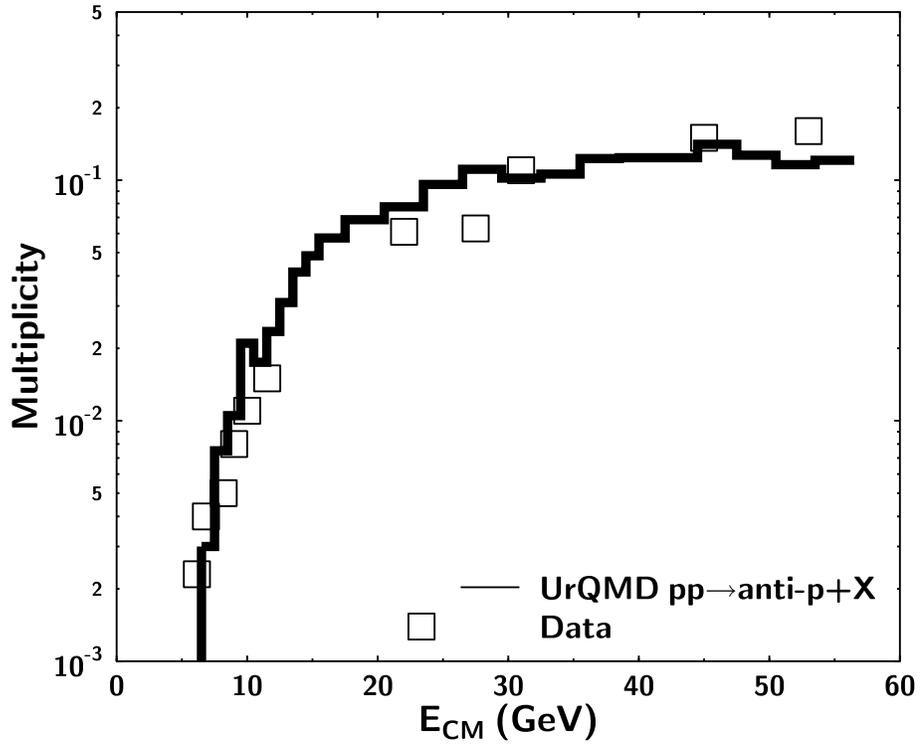,width=15cm}}
\caption{Cross section for the production of anti-protons in $pp$ as a
function of c.m. energy. The UrQMD
calculation is compared to data by \protect\cite{anti}.
}
\label{pbar}
\end{figure}

In Fig.~\ref{pbar} the cross section of $pp\rightarrow \overline p +X$ reactions
is shown. Good agreement with the data \cite{anti} is found over a
large energy range.

At higher energies, the contributions of
the different nucleonic resonances decrease and give way to the
excitation of color strings, which is  the dominant process at high
energies in our model. The total cross section above the resonance region 
is given by the CERN-HERA parameterization \cite{pdg96} as shown in 
Table \ref{tab2}.

Partonic pQCD scattering is not included into
the UrQMD model in the present version. The difference between the
total and the elastic cross-section is taken as the inelastic cross-section.

\subsection{Meson-Baryon Cross-Sections}
\label{subsec3b}

The MB cross-sections are dominated by the formation of $s$-channel
resonances, i.e. the formation of a transient state of mass
$m=\sqrt{s_{hh}}$, containing the total c.m. energy of the two incoming
hadrons. On the quark level
such a process implies that a quark from the baryon annihilates
an antiquark from the incoming meson. Below 2.2 GeV c.m. energy
intermediate resonance states get excited. The total cross-section
of these reactions are given by the expression:
\begin{eqnarray}
\label{mbbreitwig}
\sigma^{MB}_{tot}(\sqrt{s}) &=& \sum\limits_{R=\Delta,N^*}
       \langle j_B, m_B, j_{M}, m_{M} \| J_R, M_R \rangle \,
        \frac{2 S_R +1}{(2 S_B +1) (2 S_{M} +1 )}  \nonumber \\
&&\times        \frac{\pi}{p^2_{cm}}\,
        \frac{\Gamma_{R \rightarrow MB} \Gamma_{tot}}
             {(M_R - \sqrt{s})^2 + {\Gamma_{tot}^2}/{4}}\quad,
\end{eqnarray}
which depends on the total decay width $\Gamma_{\rm tot}$, on the partial
decay width $\Gamma_{R \rightarrow MB}$ and on the c.m. energy
$\sqrt{s}$. At higher energies the quark-antiquark annihilation
processes become less important. There, $t$-channel
excitations of the hadrons dominate, where the exchange of mesons and Pomeron
exchange determines the total cross-section of the MB interaction
\cite{donna}.

\begin{figure}
\centerline{\psfig{figure=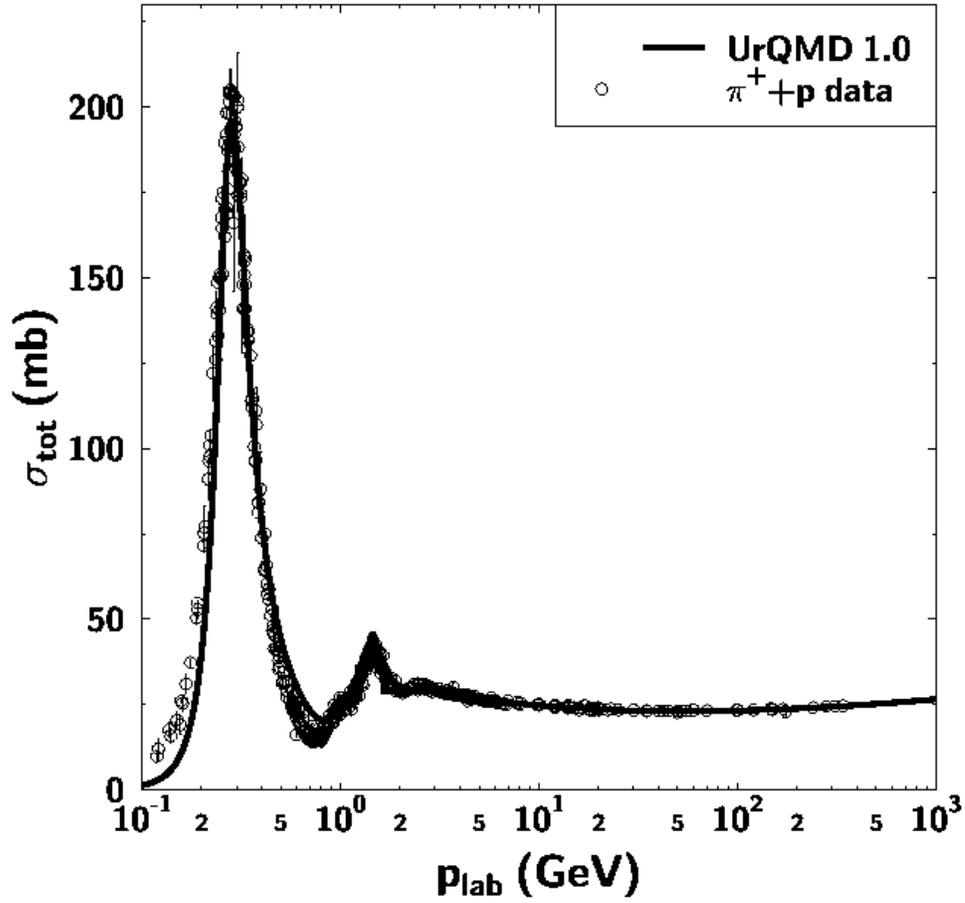,width=15cm}}
\caption{The total cross-section of $\pi^+p$ interaction vs.
laboratory momentum $p_{lab}$. Data are taken from \protect\cite{pdg96}. 
Note that the low energy s-wave $\pi^+ p$ 
scattering is not included into the UrQMD fit.
}
\label{fig5}
\end{figure}

\begin{figure}
\centerline{\psfig{figure=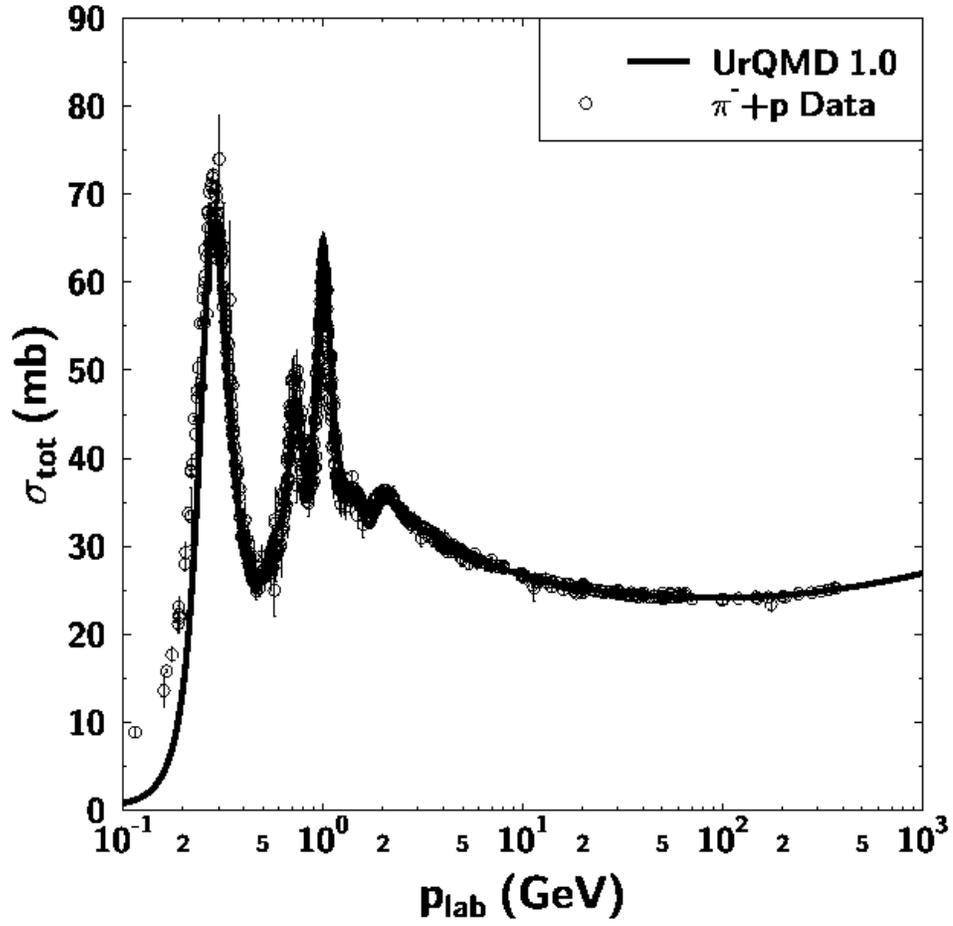,width=15cm}}
\caption{The same as Fig.~\protect\ref{fig5} but for $\pi^-p$
interaction. Data are taken from \protect\cite{pdg96}.  
Note that the low energy s-wave $\pi^- p$ scattering is not  
included into the UrQMD fit.}
\label{fig6}
\end{figure}

Figures \ref{fig5} and \ref{fig6} show the cross-section of
pion-proton reactions at different energies. In Fig.~\ref{fig5}
($\pi^++p$) one probes predominantly the creation of 
the $\Delta^{++}$ ($\Delta^{*++}$)
resonance. Note that the low energy s-wave $\pi p$ scattering is not
included into the UrQMD fit. The resonance peak at $p=1.5$~GeV/c is from
the $\Delta(1900-1950)$ resonances. In comparison Fig.~\ref{fig6}
($\pi^-+p$) depicts
many strong uncharged non-strange baryon resonances, e.g. the $\Delta^0(1232),
\Delta^{0*}(1620), \dots, N^*(1535),$ etc. The total cross-section in the
intermediate energy regime is therefore the sum of the
individual excitation modes of baryon resonances, the s-wave at lower
energies is left out.

\begin{figure}
\centerline{\psfig{figure=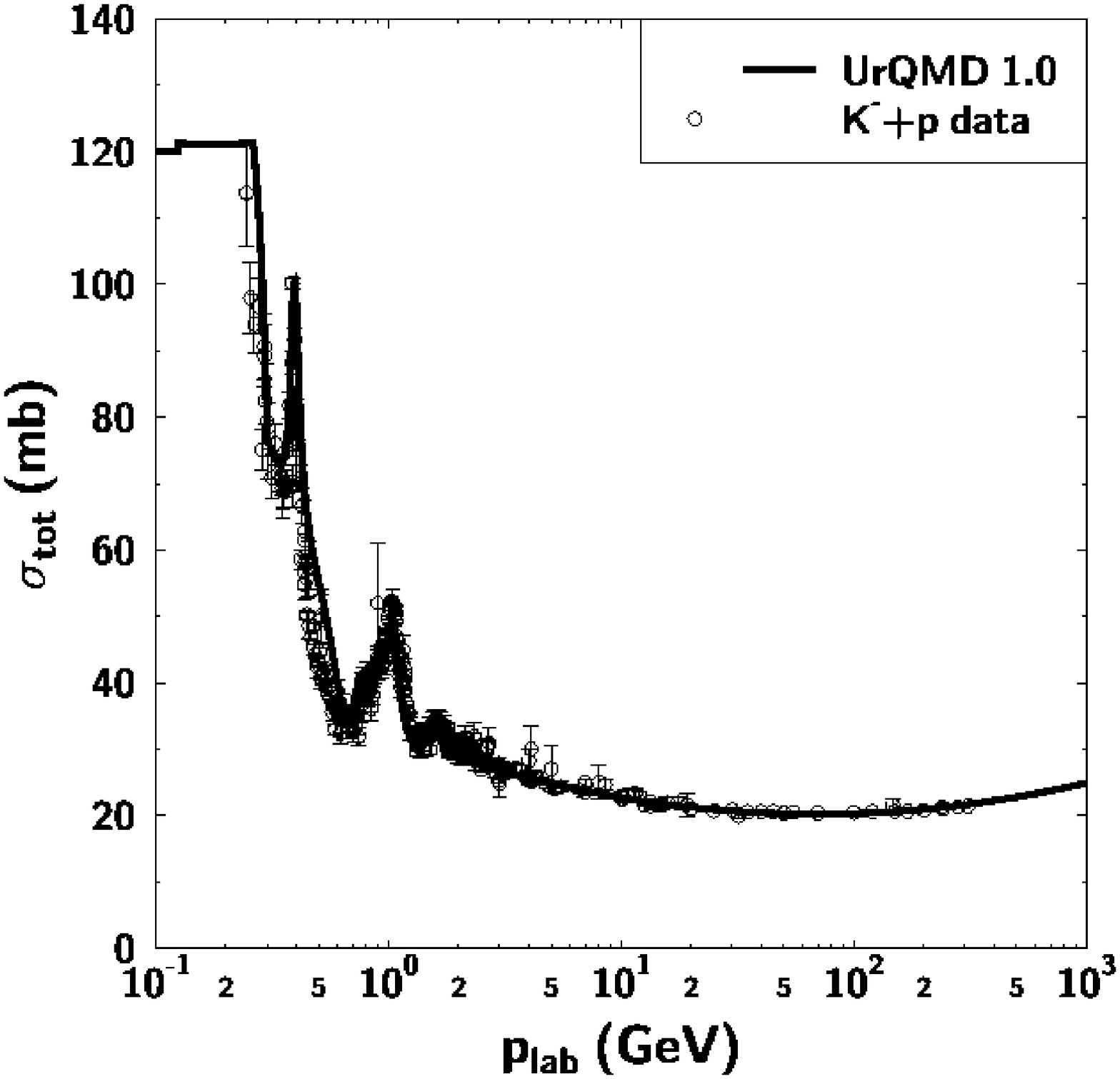,width=15cm}}
\caption{The same as Fig.~\protect\ref{fig5} but for $K^-p$
reaction. Data are taken from \protect\cite{pdg96}. }
\label{fig7}
\end{figure}

\begin{figure}
\centerline{\psfig{figure=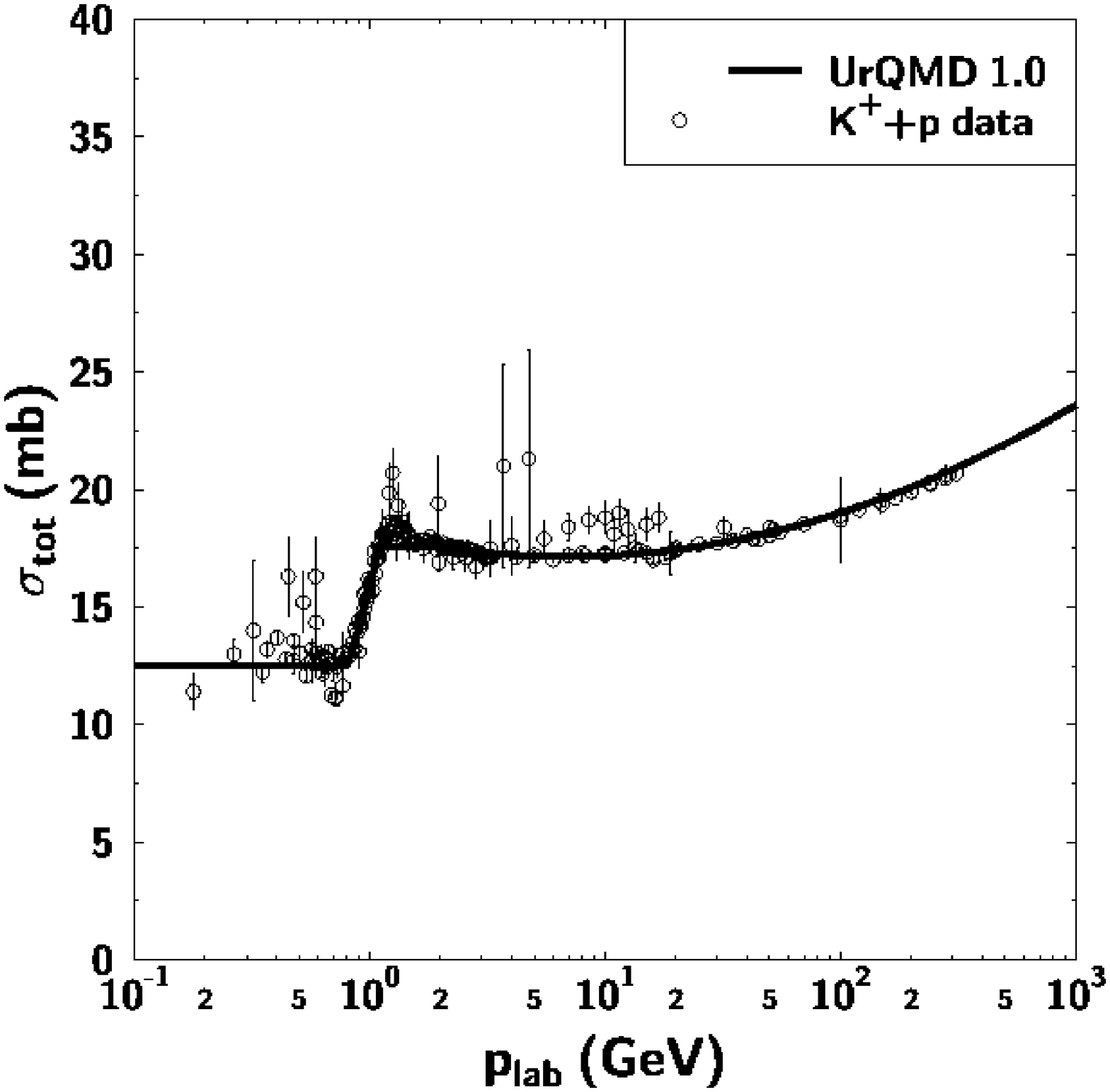,width=15cm}}
\caption{The same as Fig.~\protect\ref{fig5} but for $K^+p$
reaction. Data are taken from \protect\cite{pdg96}. }
\label{fig8}
\end{figure}

Let us now investigate collisions of strange mesons with baryons.
For $\bar{q}s$ mesons strange $s$-channel resonances can be formed on
non-strange
baryons due to the annihilation of the $\bar{q}$-quark. A comparison of
these processes from UrQMD with the experimental data is shown in
Fig.~\ref{fig7}. The formation of hyperon resonances is clearly
visible at lower energies, while the universal $t$-channel reaction
dominates the high energy tail. Fig.~\ref{fig8} shows the
cross-section of $K^+$-mesons ($u\bar{s}$) on protons. In this case,
the formation of resonances is forbidden, since the $\bar s$-quark
cannot be annihilated by non-strange baryons (For strange baryons
the formation of resonances is still possible). Here we use
only the elastic channel and the $t$-channel excitation of both
particles. The cross section at very high energies is given by the 
CERN-HERA parametrization as shown in Table \ref{tab2}. 

\subsection{Meson-Meson Cross Sections}
\label{subsec3c}

Due to the fact that the experimental preparation of meson beams and targets 
is restricted to $\pi$'s and K's,  only very little is known about MM
collisions in general.
For the description of heavy ion collisions the importance of this channel
increases with energy: at 1 A$\,$GeV beam energy we find that the
production of new hadrons (mostly pions) is only a ten percent
effect. At AGS energies (10 A$\,$GeV) the amount of mesons roughly
equals the number of incoming nucleons. Going on to the SPS (160
A$\,$GeV) the picture changes drastically: The produced particles
dominate the reactions, while the incoming nucleons have dropped to a
15\% admixture in particle density and multiplicity \cite{bleicherplb}.

To describe the total meson-meson reaction cross-sections, we make use
of the additive quark model (see below) and the principle of detailed
balance, which assumes the reversibility of the particle interactions.

\begin{figure}
\centerline{\psfig{figure=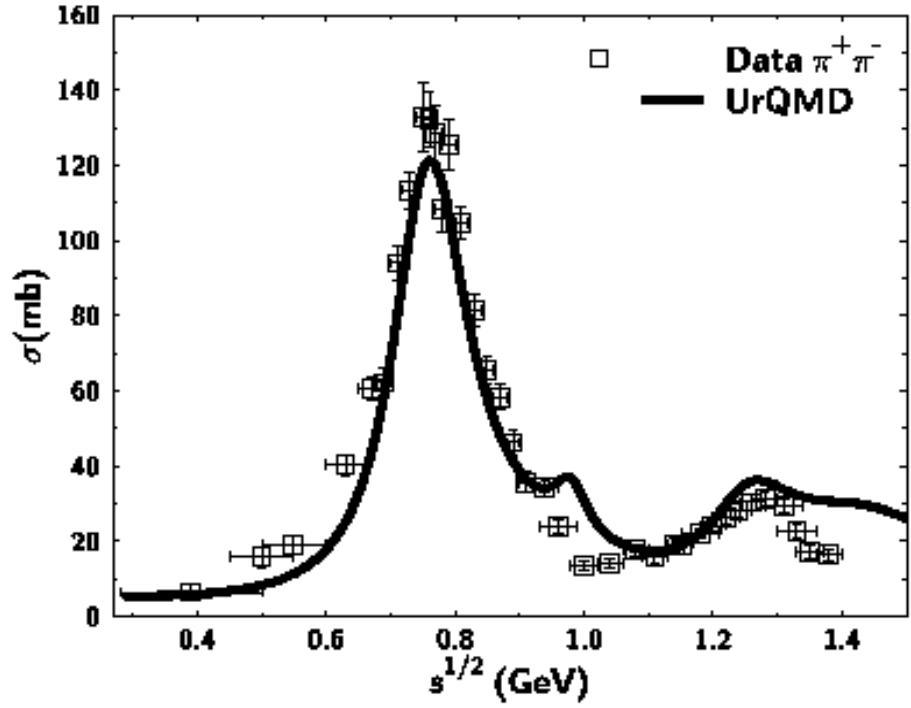,width=15cm}}
\caption{The total cross-section of $\pi^+ \pi^-$ scattering as a
function of c.m. energy $\sqrt{s}$. Data (open squares) are taken
from \protect\cite{prot73}.
}
\label{fig9}
\end{figure}

Fig.~\ref{fig9} compares the calculated cross-section of
$\pi^+ \pi^-$ scattering to experimental data \cite{prot73}.
The spectrum is dominated by the formation of the $\rho$ with a mass
of 770~MeV, the other two small peaks belong to the $f_0(970)$ and
$f_2(1270)$ resonances. The $f_0(970)$ resonance is not visible in the
data since the experimental analysis of meson-meson scattering is model
dependent.

\begin{figure}
\centerline{\psfig{figure=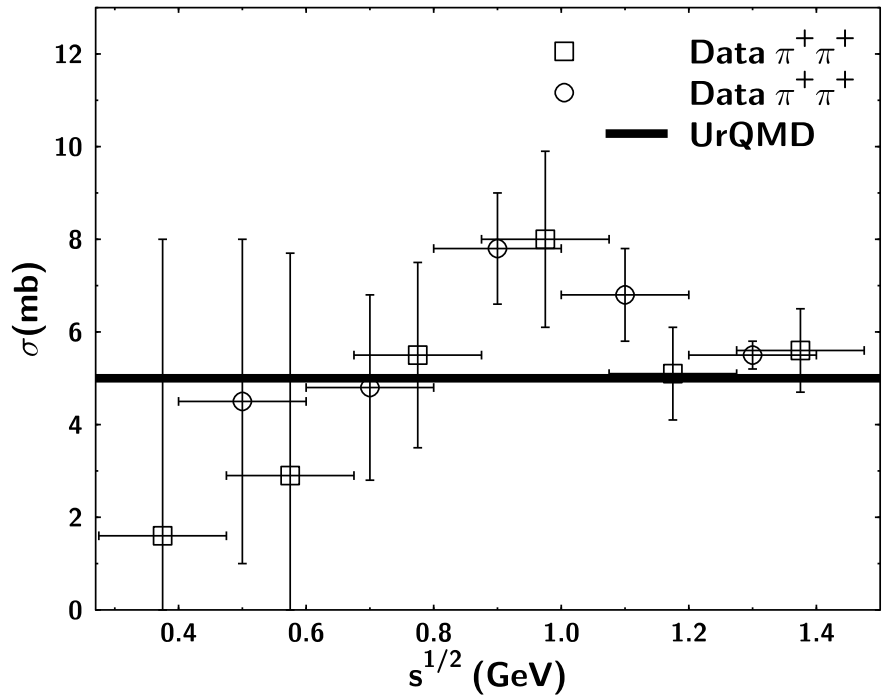,width=15cm}}
\caption{The same as Fig.~\protect\ref{fig9} but for $\pi^+ \pi^+$
scattering. Data are taken from \protect\cite{coh73} (open squares)
and from \protect\cite{dur73} (open circles).
}
\label{fig10}
\end{figure}

Figure~\ref{fig10} shows the implemented elastic $\pi^+ \pi^+$
cross-section, which remains constant in the whole energy range where
data are available \cite{coh73,dur73}.

\begin{figure}
\centerline{\psfig{figure=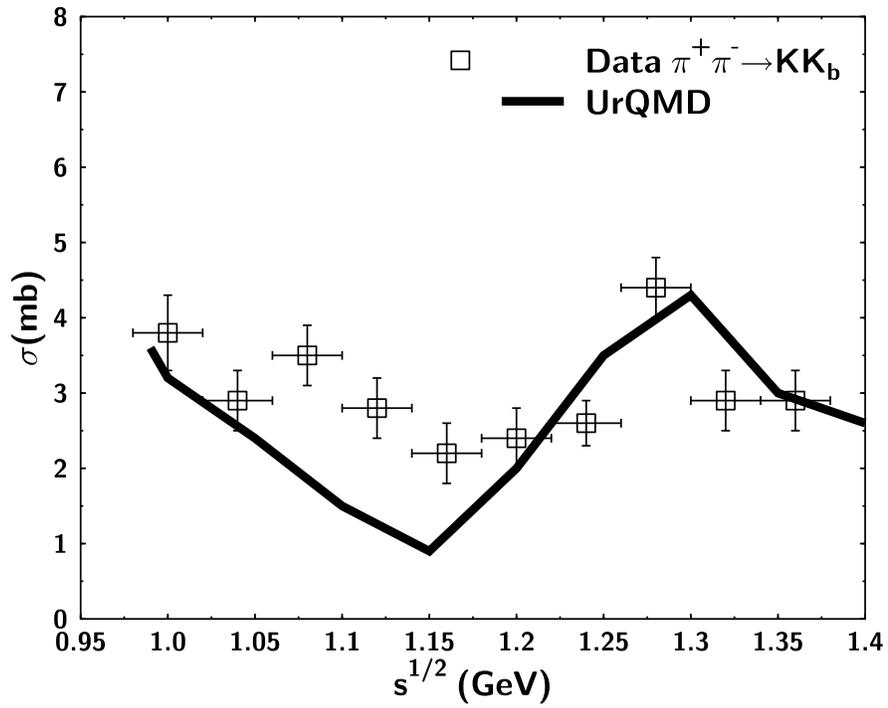,width=15cm}}
\caption{Cross-section of the reaction $\pi^+ \pi^- \rightarrow K
\bar K$ as a function of $\sqrt{s}$. Data (open squares) are taken
from \protect\cite{prot73}.
}
\label{fig11}
\end{figure}

Strangeness production in the meson-meson
channel is possible, e.g. via the reaction
$\pi^+\pi^-\rightarrow K\bar K$ as shown in Fig.~\ref{fig11}.

\begin{figure}
\centerline{\psfig{figure=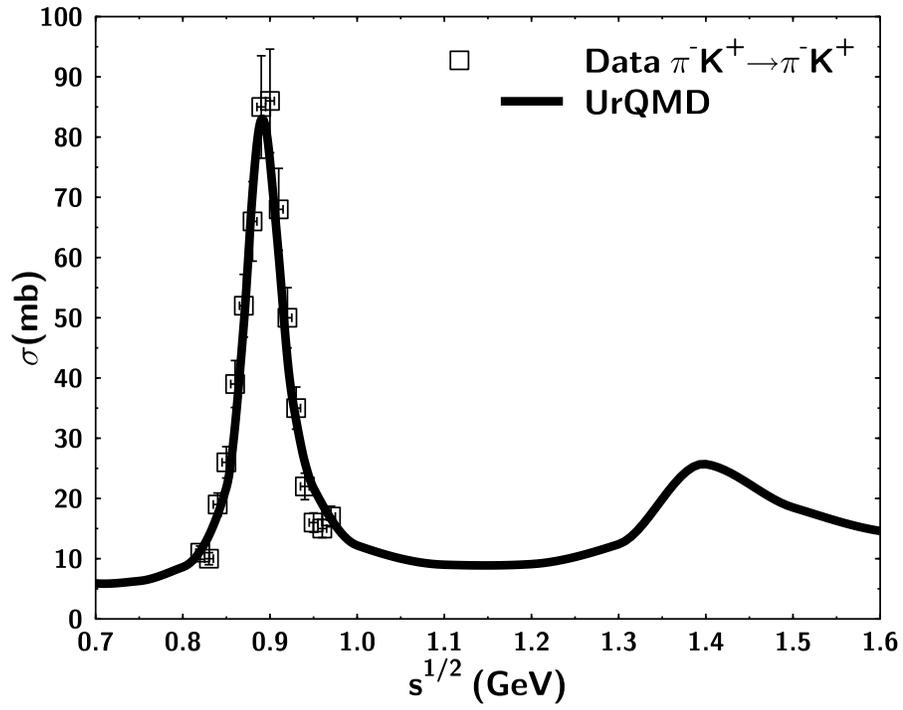,width=15cm}}
\caption{Cross-section of $\pi^- K^+$ scattering vs. $\sqrt{s}$.
Data (open squares) are taken from \protect\cite{mat74}.
}
\label{fig12}
\end{figure}

\begin{figure}
\centerline{\psfig{figure=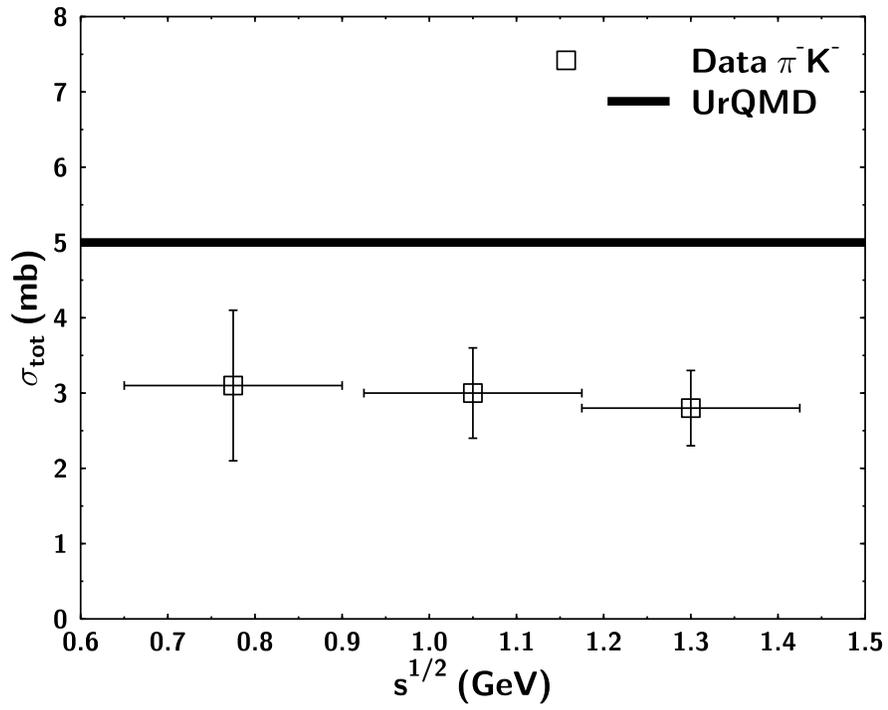,width=15cm}}
\caption{The same as Fig.~\protect\ref{fig12} but for $\pi^- K^-$
reaction. Data (open squares) are taken from \protect\cite{ling73}.
}
\label{fig13}
\end{figure}

Rescattering of strange mesons is implemented via the resonance
formation (cf. Fig.~\ref{fig12} dominated by the $K^*$ resonance
\cite{mat74}), or elastically as depicted in Fig.~\ref{fig13} \cite{ling73}.

\begin{figure}
\centerline{\psfig{figure=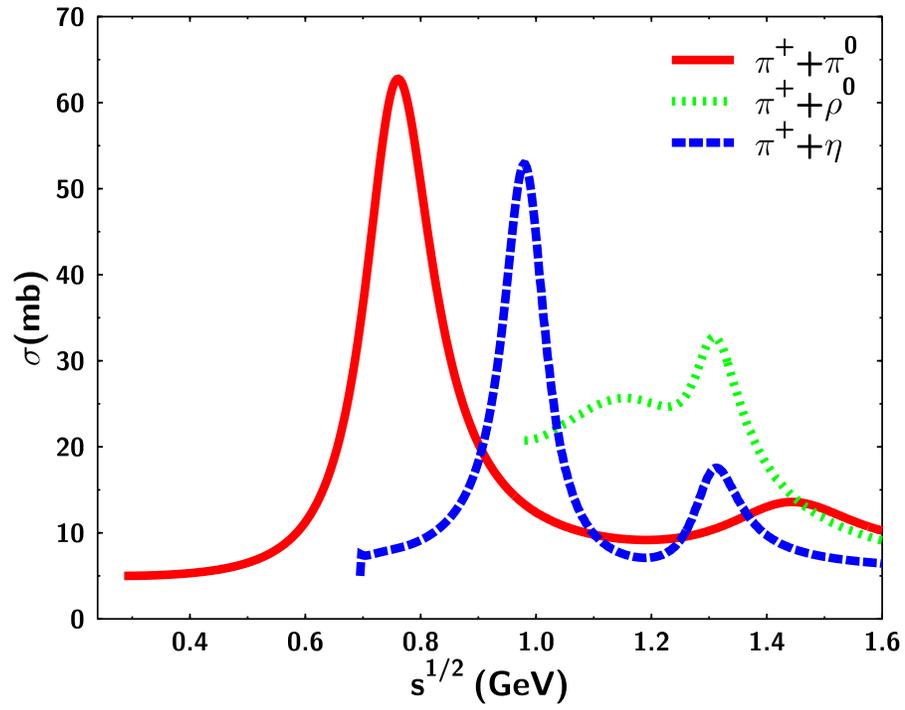,width=15cm}}
\caption{Cross-sections of $\pi^+ \pi^0$ (solid curve), $\pi^+ \rho^0$
(dotted curve) and $\pi^+ \eta$ (dashed curve) as functions of
$\sqrt{s}$.
}
\label{fig14}
\end{figure}

Finally, we predict the cross-sections of $\pi^+\pi^0$, $\pi^+\rho^0$
and $\pi^+\eta$-reactions
(Fig.~\ref{fig14}) which are of utmost importance for the production
of thermal photons and dileptons. At higher energies other meson resonances
can be
formed. To model MM interactions above the resonance region $\sqrt s >
1.7$~GeV
we use the rescaled total $\pi p$ cross section:
\be
\sigma_{\rm tot}^{MM}(\sqrt s > 1.7\mbox{ GeV})=\sigma_{\rm tot}^{\pi
p}(\sqrt s)
\frac{\sigma_{\rm AQM}^{MM}}{\sigma_{\rm AQM}^{\pi p}}\quad.
\ee
This is justified, since at high energies the total cross section is given
by quark
counting. In the energy region from $\sqrt s > 1.7$~GeV to $\sqrt s
<6$~GeV $s$-channel interactions are taken into account, while from
$\sqrt s > 3$~GeV on $t$-channel excitation of both mesons becomes the
MM interaction process of increasing importance in the model.

The cross section for high energetic reactions are taken from the
AQM-rescaled $\pi^+ p$ paramatrization by the CERN-HERA group 
(see Table \ref{tab2}).

\subsection{Antibaryon$-$Baryon Cross-Sections}
\label{subsec3d}

The physics of baryon$-$antibaryon interactions has been an area of
much theoretical and experimental activity for a rather long period.
It is well-known that at energies $p_{\rm lab} \leq 100\,$GeV/$c$ an
important contribution to the total interaction cross-section comes
from the process of annihilation, where  only mesons are left in the
final state. Though the earlier experiments on
$\bar p p -$annihilation revealed a number of differences from the
non-annihilation channels, it is not clearly understood whether these
differences arose simply due to the kinematic restrictions on the
available phase space, or whether they are related to dynamical differences
between the non-annihilation and annihilation mechanisms.
The experimental results obtained in \cite{zabr95} by comparison
of $pp$ with non-annihilation $\bar{p}p$ interactions at 32 GeV/$c$
support the conclusion of equivalence of $pp$ and non-annihilation
$\bar{p}p$ interaction processes.

Still, the nature of the baryon annihilation is subject
to theoretical discussions.
In the framework of the quark models based on topological $1/N$
expansion, the annihilation process is associated with the annihilation
of string junctions, i.e. the point where strings are connected, such
that the baryons have a $Y$-form. In this case three $q\bar{q}$ strings are
formed. The theory also  allows for diagrams where the
string junctions and one or two of the valence quarks can annihilate,
corresponding
to the formation of two strings or one string.
Other theories consider an annihilation mechanism without invoking the
string junction hypothesis. This intriguing question has yet to
be clarified (for review see, e.g. \cite{dover92}, and references
herein).

\begin{figure}
\centerline{\psfig{figure=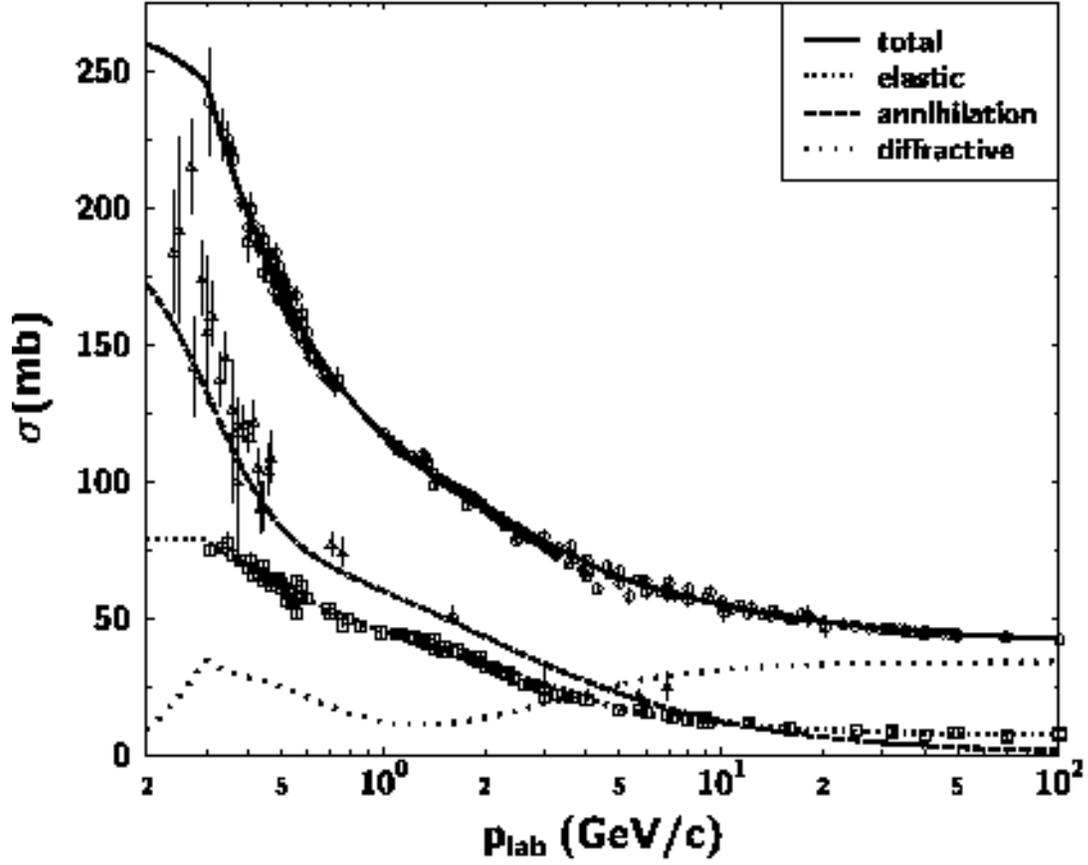,width=15cm}}
\caption{The $\bar p p$ cross-section as compared to the experimental
data on total (open circles), elastic (open squares), and annihilation
(open triangles) cross-sections. Data are taken from \protect\cite{pdg96}. 
The diffractive cross-section is assumed to be a
difference between the total cross-section and the sum of the elastic
and annihilation cross-section.
}
\label{fig15}
\end{figure}

To avoid the difficulties attributed to these theoretical approaches,
UrQMD is adjusted to known experimental data. The
total  $\bar p p$ cross-section is shown in Fig.~\ref{fig15}, as
well as the annihilation and the elastic cross-sections. The UrQMD
parameterizations depicted by lines are taken from Koch and Dover \cite{koch}:
\be
\sigma _{\rm ann}^{\bar p p} \,=\, \sigma _{0}^{N} \, \frac {s_0}{s}
\left [
\frac {{\rm A}^2 s_0}{(s - s_0)^2 + {\rm A}^2 s_0} + {\rm B} \right ]
\quad ,
\ee
with $\sigma _{0}^{N} = 120$ mb, $s_0 = 4 m_{N}^2$,
A = 50 MeV and B = 0.6.
The $\bar n p$ cross-section does not differ significantly from the
$\bar p p$ cross-section \cite{elioff62}, hence they are set equal in
the UrQMD model.

At higher energies, CERN-HERA parameterizations \cite{pdg96} are used for
the total and elastic channel:
\be
\sigma^{\bar p p}_{tot,el}(p)\,=\,{\rm A} + {\rm B}\, p^n +
{\rm C}\,{\rm ln}^2(p) + {\rm D}\,{\rm ln}(p)\quad ,
\ee
with the laboratory momentum $p$ in GeV/$c$ and the cross-section
$\sigma$ in mb. The parameters of the fit are listed in
Table~\ref{tab2}.

Below $p_{lab}<5$~GeV/c the following parameterization is used:
\be
\sigma_{\rm tot}(p) = \left\{ \begin{array}{ll}
75.0+43.1 p^{-1} + 2.6 p^{-2} -3.9 p &\quad ;\quad 0.3<p<5\\
271.6 \exp{(-1.1\, p^2)} & \quad;\quad p< 0.3
\end{array} \right.
\ee

\be
\sigma_{\rm el}(p) = \left\{ \begin{array}{ll}
31.6+18.3 p^{-1} -1.1 p^{-2} - 3.8 p &\quad ;\quad 0.3<p< 5\\
78.6 &\quad ;\quad p< 0.3\;
\end{array} \right.
\ee

The sum of annihilation and elastic cross-sections do not
yield the total cross-section:
\be
 \Delta\sigma = \sigma_{\rm tot}-\sigma_{\rm el}-\sigma_{\rm ann}
\ee
In UrQMD this difference $\Delta\sigma$ is interpreted as the diffractive
cross-section which describes the excitation at least one of the collision
particles to a resonance or to a string via Pomeron exchange.

\begin{figure}
\centerline{\psfig{figure=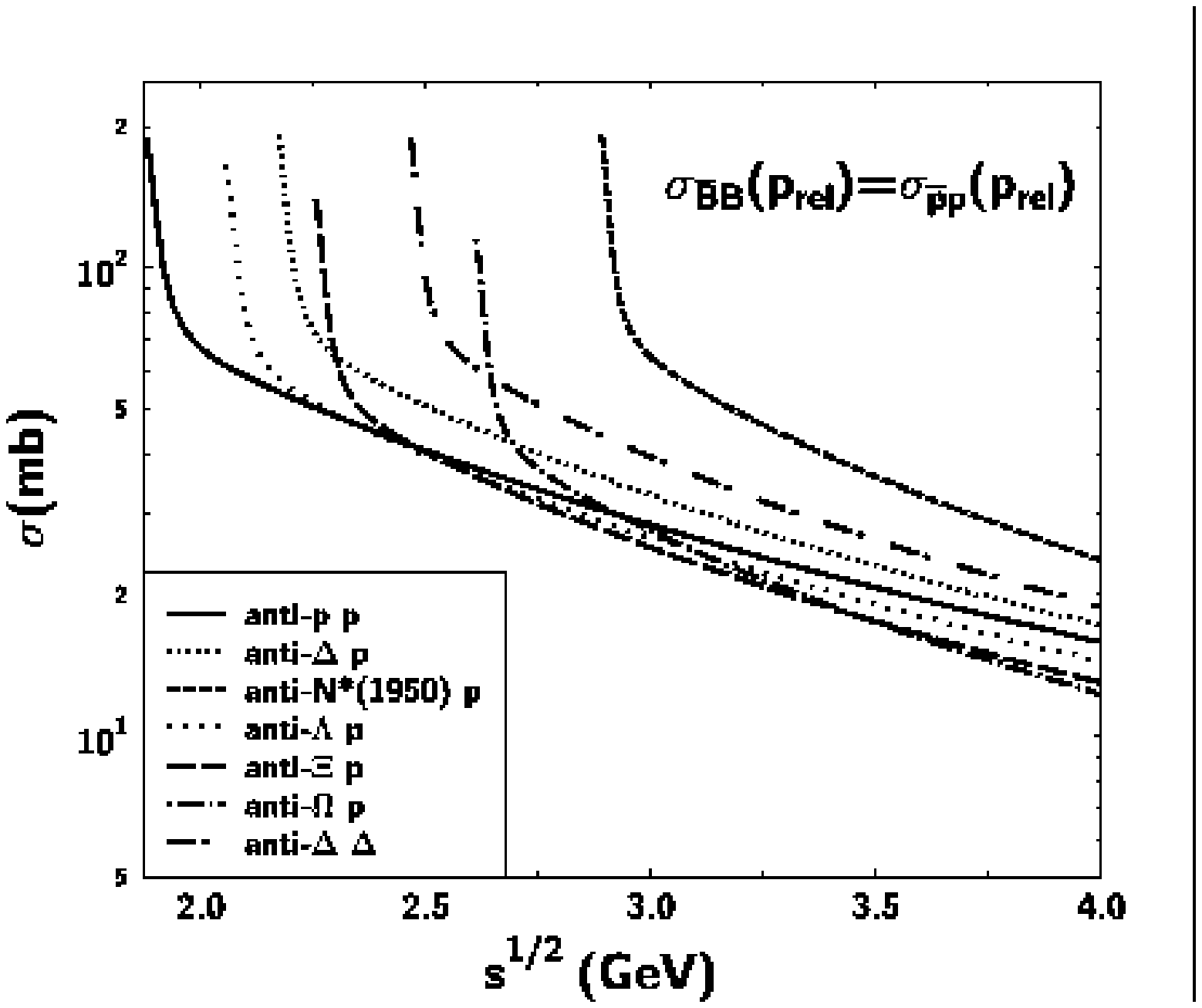,width=15cm}}
\caption{Extrapolation of the $\bar p p$ cross-section towards unknown
antibaryon$-$baryon reactions. Here the cross-section of the
antibaryon$-$baryon interaction is equal to the $\bar p p$ cross-section
at the same relative momentum. In the UrQMD model we take the
antibaryon$-$baryon cross section equal to the $\bar p p$ cross-section 
at the same center-of-mass energy.
}
\label{fig16}
\end{figure}

The annihilation of baryon$-$antibaryon pairs proceeds in the
UrQMD model according to {\it rearrangement\/} diagrams. Here
the formation of two $q \bar q$-strings of equal energies in
the c.m. system is assumed while the remaining constituent
quarks are rearranged into newly produced hadrons.
The generalization of the $\bar{p}p$ cross-section towards all
possible antibaryon$-$baryon collisions can be done in different
ways:
\begin{enumerate}
\item{The anti-baryon baryon cross-section at a given c.m. energy
$\sqrt{s}$ is equal to the $\bar{p}p$ annihilation cross-section at
the same $\sqrt{s}$:
\be
\sigma_{\bar{B}B}|_{\sqrt{s}}=\sigma_{\bar{p}p}|_{\sqrt{s}}
\ee
}
\item{The anti-baryon baryon cross-section at a given relative momentum
$p_{rel}$ is equal to the $\bar{p}p$ annihilation cross-section at
the same $p_{rel}$ (Fig.~\ref{fig16}):
\be
\sigma_{\bar{B}B}|_{p_{rel}}=\sigma_{\bar{p}p}|_{p_{rel}}
\ee
}
\end{enumerate}

In the UrQMD we have chosen the first parameterization. 
Since $\sigma_{\rm ann} \propto s^{-1/2}$, the
annihilation cross-section drops rapidly with rising particle
mass. The different treatment this
cross-section can lead to systematic shifts in the antibaryon
distributions for massive systems. Therefore, anti-baryon production off
nuclei may be used to solve this ambiguity.

\subsection{The Additive Quark Model (AQM)}

\begin{figure}
\centerline{\psfig{figure=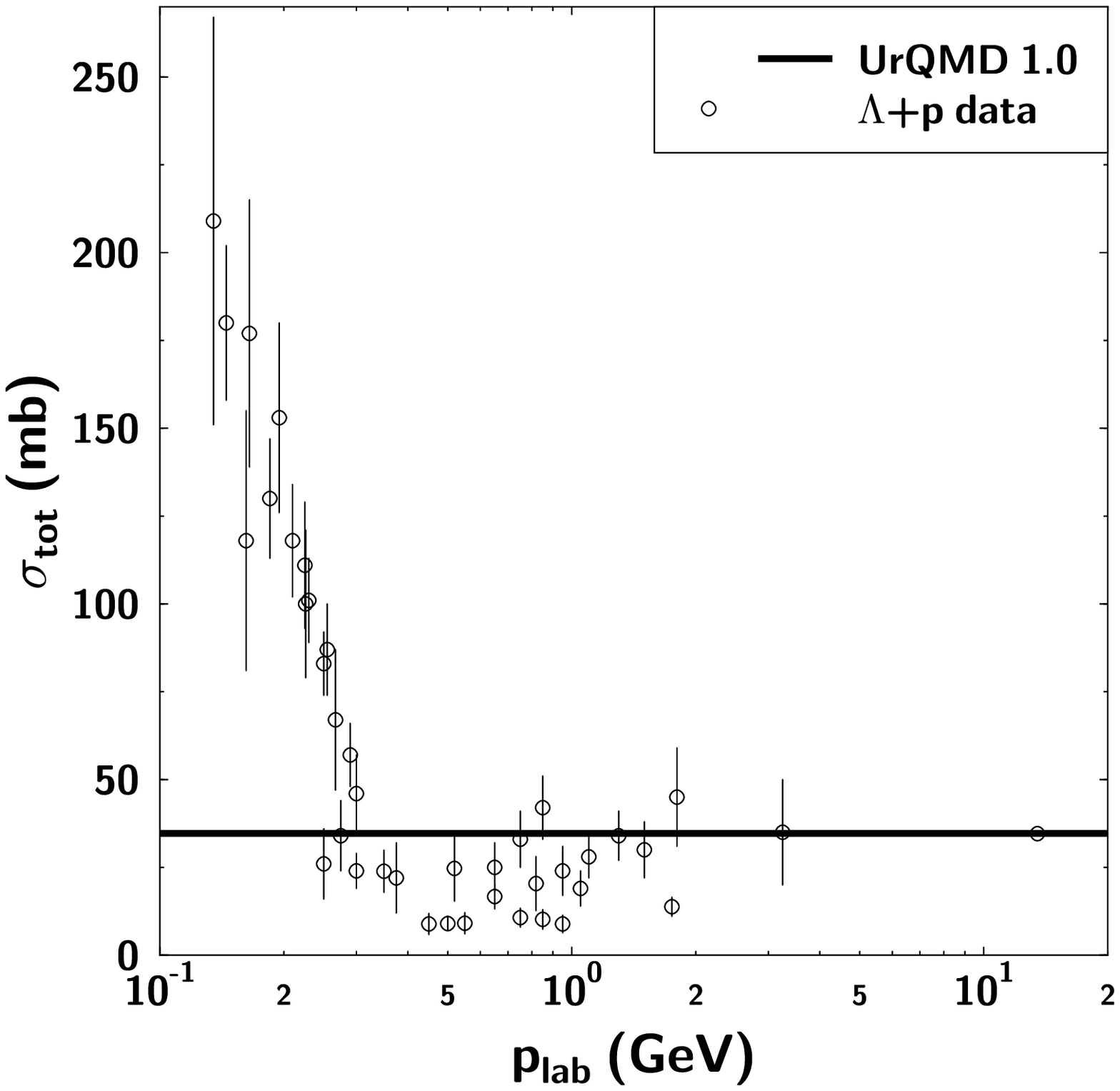,width=15cm}}
\caption{Total $\Lambda$-p cross-section vs. laboratory
momentum of the $\Lambda$. The UrQMD results are given by the Additive Quark
Model. Data are taken from \protect\cite{pdg96}. 
There seems to be indication for a resonance at $\sqrt s =2.1$~GeV. This
could be a 6q molecule or a di-baryon with $s=1$.
}
\label{fig17}
\end{figure}

Unknown cross-sections are calculated on the basis of the
Additive Quark Model (AQM) \cite{clos79,perk87}, which assumes the
existence of dressed valence quarks, interacting very weakly inside of
the hadron. At the phenomenological level, the AQM gives a correct
quantitative and qualitative description of, e.g., the asymmetry of
c.m. spectra of secondaries in meson$-$nucleon and photon$-$nucleon
reactions. The AQM has predicted also the important role of resonances for
the multiple production of particles in hadronic interactions. To
apply the AQM to calculations of heavy ion reactions, one
needs to know the cross-sections of the quark
interactions, which can be evaluated from the quark masses. Then, the
unknown total cross-section of the high energy  reaction
can be calculated assuming a 40\% reduced $s$-quark
cross-section (compared to that of $u$- and $d$-quark). The elastic
cross-section is derived from Regge theory \cite{goul83}:
\be
\sigma_{\rm elastic} = 0.039 \sigma_{\rm total}^{\frac{3}{2}}
\, [{\rm mb}]\quad,
\ee
where
\be
\sigma_{\rm total} = 40\left(\frac{2}{3}\right)^{m_1+m_2}
\left(1-0.4\frac{s_1}{3-m_1}\right) \left(1-0.4\frac{s_2}{3-m_2}
\right) \, [{\rm mb}]\quad,
\ee
where $m_i =1 (0)$ for particle $i$ being a meson (a baryon) and $s_i$
is the number of strange quarks in particle $i$. This
formula results from the high energy reactions, therefore there is no
difference
between antiparticles and particles. For BB-reactions
no additional energy dependence is employed in collisions involving
strange baryons.  Non-strange baryon cross-sections are not treated via the
Additive Quark
Model, they have an explicit energy dependence in line with experimental
data.
The MB and MM cross-sections are rescaled via:
\be
\sigma_{X_1X_2}(\sqrt s)=\frac{\sigma_{\pi N}(\sqrt s)}{\sigma_{\pi N}^{\rm
AQM}}\,
\sigma_{X_1X_2}^{\rm AQM}\quad.
\ee
The hyperon-nucleon cross-section, which is taken in the UrQMD model from
the Additive Quark Model, is in good agreement with the data above
$p_{\rm lab}=300$~MeV (cf. Fig.~\ref{fig17}). The total cross-sections
calculated for
baryon-baryon, meson-baryon and meson-meson interactions are listed
in Tables~\ref{tab3}$-$\ref{tab5}.

\subsection{Color Fluctuations, Color Opacity and Color Transparency}

Quantum chromodynamics has important applications of the dynamical
role of color degrees of freedom to the strong interactions at
ultrarelativistic energies (for a review, see \cite{fran94,baym96}
and references therein). The theory is presented in
Ref.~\cite{fran94} in detail, here we just sketch the
main ideas of color optics and coherent phenomena in high energy
physics. Hadrons are dynamical objects which come in Fock space
 of configurations of very different
spatial sizes. At high energies, incident hadronic quark$-$gluon
configurations
can be considered frozen as a result of Lorentz time dilation. Due to the
long coherence length at such high energies one can apply geometrical color
optics. Small objects produced in hard processes with high
momentum transfer $Q^2$ then  have reduced interaction cross sections. In
processes with
moderate $Q^2$ such compact objects, which are a coherent superposition
of eigenstates of the QCD Hamiltonian, should gain size. When the
quark$-$gluon configuration is large, it will lead to an increased
interaction cross-section of the hadron with the nuclear medium.

Therefore, the fluctuations of the hadron's spatial extent give rise to the
color transparency and color opacity phenomena: When a small
object is produced, it interacts only very weakly with other hadrons due to
color screening. Moreover, since - at sufficiently high energies - the
small-sized configuration of this object is frozen, the nuclear medium
appears to be transparent for such hadrons (color transparency). In
contrast, hadronic configurations which are larger than average
interact with larger cross-section, giving rise to color opacity.

Nucleus-nucleus collisions provide a tool to investigate the effect of
color transparency, for instance, in the production of leading
nucleons. On the other hand, the complementary color opacity effect,
i.e. large-sized configurations, can cause stronger stopping and
significant fluctuations in the transverse energy of secondaries in
central reactions.

\begin{figure}
\centerline{\psfig{figure=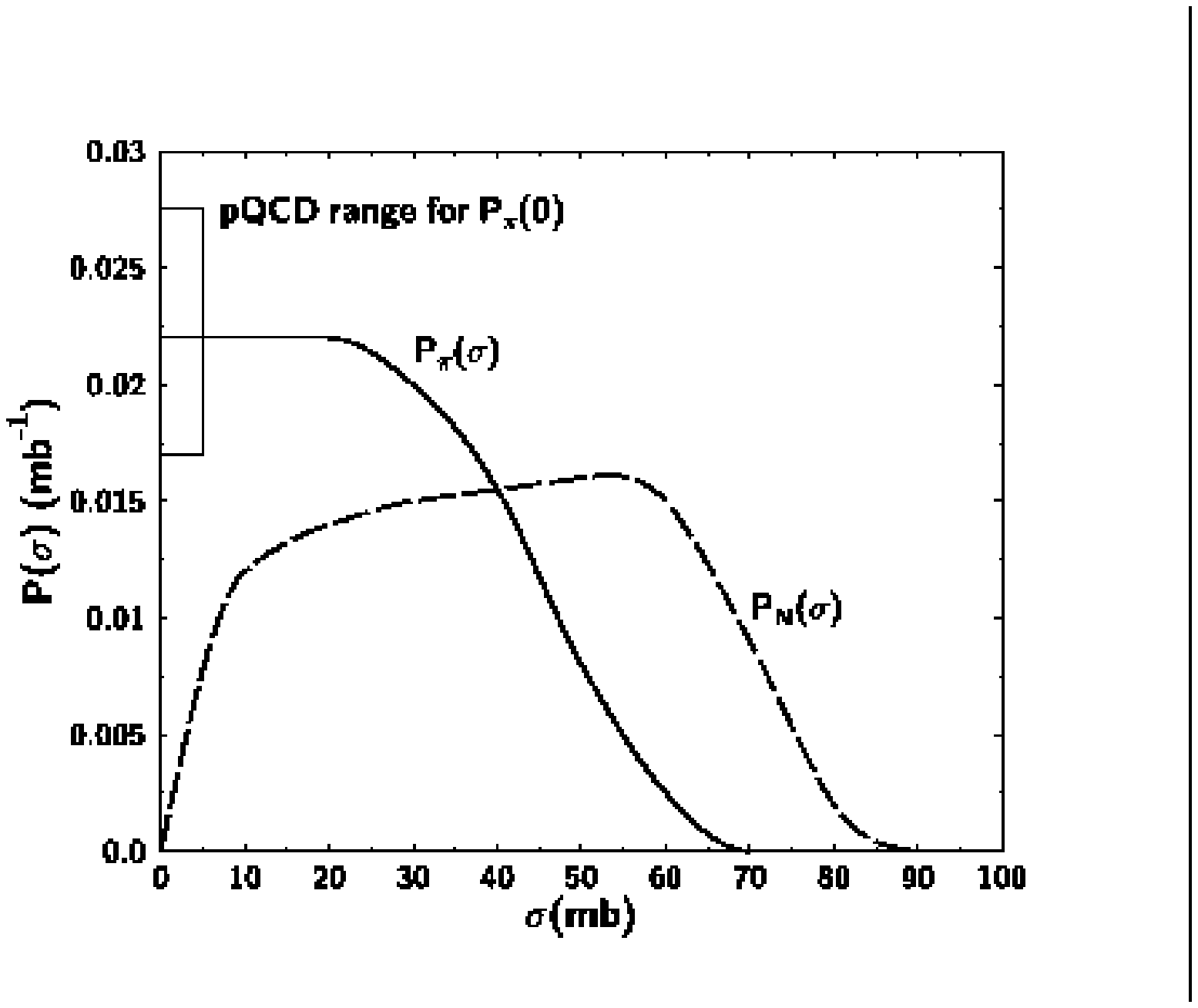,width=15cm}}
\caption{Probability distribution of nucleon (dashed curve) and pion
(solid curve) cross-sections as predicted by \protect \cite{fran94}.
}
\label{fig18}
\end{figure}

A first step to investigate these QCD effects within a microscopic
transport model is made by incorporating the color fluctuations in the
elementary hadron$-$hadron reactions in the UrQMD model. Thus,
one needs to know the
probability $P(\sigma)$ that a given configuration interacts with a
nucleon with a total cross-section $\sigma$.
It is convenient to consider moments of the distribution:
\bea
\langle \sigma^0 \rangle &=&\int{\rm d}\sigma \bar{\sigma}^0 P(\sigma) 
        =1\quad, \\
\langle \sigma^1 \rangle &=&\int{\rm d}\sigma \bar{\sigma}^1 P(\sigma)
=\bar{\sigma}\quad, {\rm etc.}\quad,
\eea
where $\bar{\sigma}$ denotes the average cross-section.
The second moment $<\sigma^2>$ can be determined from the diffractive
dissociation experiments. In addition, further information can be
obtained from QCD, which implies \cite{fran94}:
\be
P(\sigma)\propto\sigma^{N_q-2}\quad,
\ee
for $\sigma\rightarrow 0$, where $N_q$ is the number of valence
quarks. Thus, for the nucleon and the pion distributions it follows for
$\sigma\rightarrow 0$:
\bea
P_{N}(\sigma)  &\propto& \sigma \quad,\\
P_{\pi}(\sigma)&\propto& {\rm constant}\quad.
\eea
From these arguments, $P(\sigma)$ can be construct.
Fig.~\ref{fig18} shows the resulting broad $P(\sigma)$
distribution for proton projectiles and the even broader one for the
pions \cite{fran94}.

\begin{figure}
\centerline{\psfig{figure=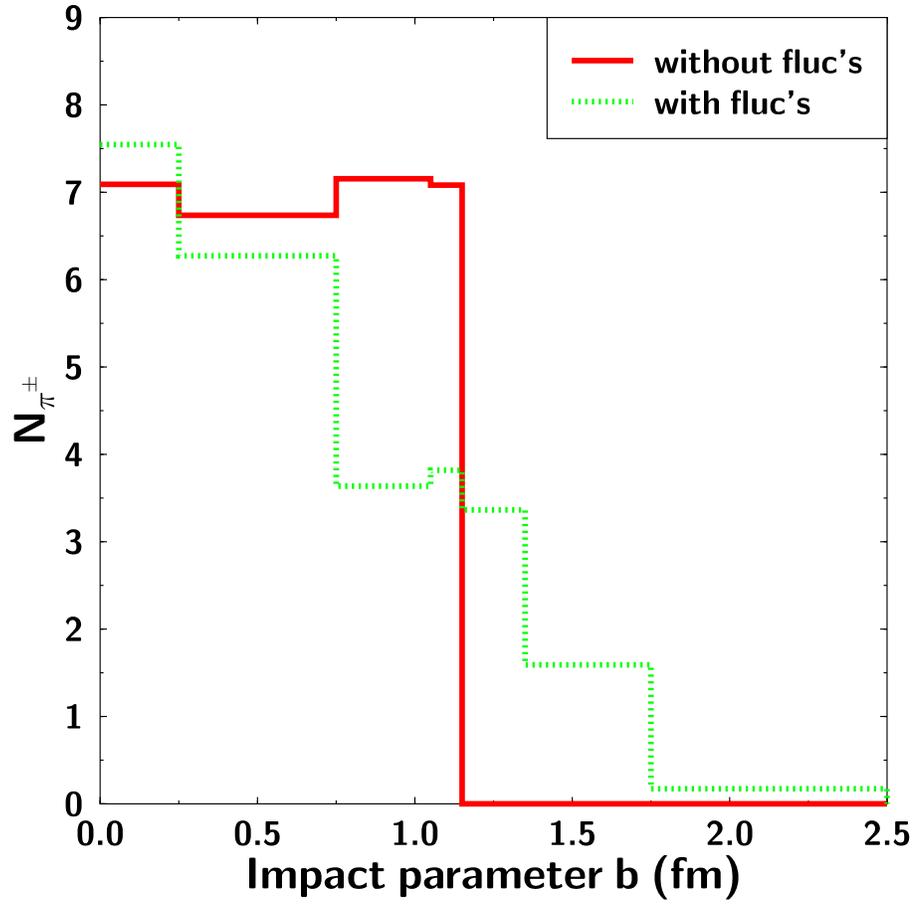,width=15cm}}
\caption{Charged pion multiplicity in $pp$ collisions at
$\sqrt s = 27$~GeV with (dotted-) and without (solid-line histogram)
color fluctuations for different impact parameters.
}
\label{fig19}
\end{figure}

The effect of the color fluctuations on proton-proton interactions at
different impact parameters, $b$, is shown in Fig.~\ref{fig19} for the
UrQMD model.
The charged pion multiplicity distribution decreases
monotonically with rising $b$. $N_{\pi}(b)$ has a non-vanishing tail
for $b \geq 1.1\,$fm, in contrast to the abrupt geometrical edge of the
distribution
as calculated in the static geometric models without color fluctuations.

\section{The Reaction Channels}
\label{sec4}


\subsection{Resonances}
\label{res_sec}

The production and decay of resonances is the most important h-h reaction
channel below $\sqrt{s}=5$~GeV for BB and 3~GeV for MM and MB reactions.
Baryon resonances are produced in two
different ways, namely 
\begin{itemize}
\item[\bf i)] {\it hard production\/}:
N+N$\rightarrow \Delta$N,\ $\Delta\Delta$,\ N$^*$N, etc.
\item[\bf ii)] {\it soft production\/}: $\pi^-+$p$\rightarrow
\Delta^0$,\ K$^-$+p$\rightarrow \Lambda^*$...
\end{itemize}

The formation of $s$-channel resonances is fitted to measured data,
e.g. in the reaction $A + C \rightarrow D + E$ we use the general form
\be
\sigma_{tot}^{BB}(\sqrt{s}) \propto (2S_D + 1) (2S_E + 1) \frac{
\langle p_{D,E} \rangle }{ \langle p_{A,C} \rangle }\, 
\frac{1}{s}\, |{\cal M}(m_D,m_E)|^2 \quad,
\ee
with the spins of the particles, $S_i$, momenta of the pairs of
particles, $<p_{i,j}>$, in the two-particle rest frame, and the matrix
element $|{\cal M}(m_D,m_E)|^2$, 
which here depends only on the masses of the
outgoing hadrons, $m_i$.

\begin{figure}
\centerline{\psfig{figure=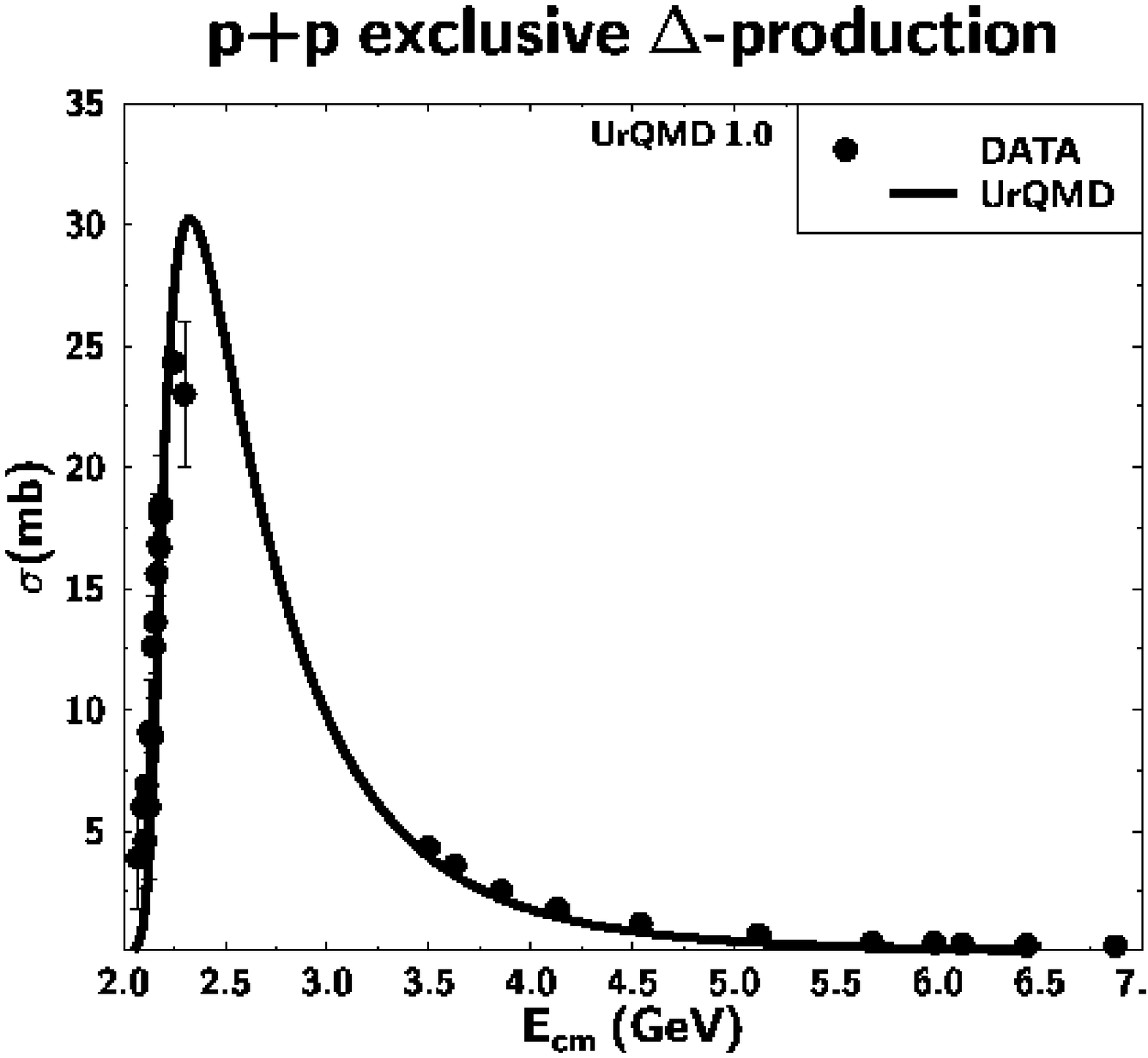,width=15cm}}
\caption{UrQMD fit for the exclusive $\Delta_{1232}$ 
production in proton-proton
reactions  compared to data \protect\cite{flaminio}.
}
\label{pp-nd1232}
\end{figure}

There are six channels of the excitation of
non-strange resonances in the UrQMD model, namely $NN \rightarrow N
\Delta_{1232}, N N^{\ast}, N \Delta^{\ast}, \Delta_{1232}
\Delta_{1232}, \Delta_{1232} N^{\ast},\ {\rm and} \ \Delta_{1232}
\Delta^{\ast}$. 
The $\Delta_{1232}$ is explicitly listed,
whereas higher excitations of the $\Delta$ resonance
have been denoted as $\Delta^{\ast}$.
For each of these 6 channels specific assumptions are made
with respect to the form of the matrix element ${\cal M}$, and the free
parameters are adjusted to the available experimental data, when
available:

\begin{enumerate}
\item $N N \to N \Delta_{1232}$ excitation:
\begin{equation}
| {\cal M}(\sqrt{s},m_3,m_4) |^2 \,=\, A\, \frac{m_\Delta^2 \Gamma_\Delta^2}
{\left( ( \sqrt{s})^2 - m_\Delta^2 \right)^2 + m_\Delta^2 \Gamma_\Delta^2}
\quad,
\end{equation}
with $m_\Delta=1.232$ GeV, $\Gamma_\Delta=0.115$ GeV and $A=40000$. Note that
this form of the matrix element has been adjusted to fit the data shown in
figure \ref{pp-nd1232}.
\item $N N \to N N^*$, $N N \to N \Delta^*$, 
$N N \to \Delta_{1232} N^*$ and $N N \to \Delta_{1232} \Delta^*$ excitation:
\begin{equation}
| {\cal M}(m_3,m_4) |^2 \,=\, A\, \frac{1}
{(m_4-m_3)^2 \, (m_4+m_3)^2}
\quad,
\end{equation}
with $A=6.3$ for $N N \to N N^*$, $A=12$ for $N N \to N \Delta^*$
and $A=3.5$ for  $N N \to \Delta_{1232} N^*$. In the case of
$N N \to \Delta_{1232} \Delta^*$ there are insufficient data available,
therefore we use the same matrix element and parameters as in the
case of $N N \to \Delta_{1232} N^*$.
Since $m_3 \ne m_4$ is valid for all above cases, 
the matrix element cannot diverge.
\item $N N \to \Delta \Delta$ excitation:
\begin{equation}
| {\cal M}(m_3,m_4) |^2 \,=\, A
\quad,
\end{equation}
with $A=2.8$.
\end{enumerate}

Figure~\ref{pp-nd1232} shows  the fit of the UrQMD
$ p p \to N \Delta_{1232}$ cross section to experimental measurements
\cite{flaminio}. The measurements refer to the $\Delta^++ n$ 
exit channel and have been rescaled to match the full isospin-summed
cross section. In the case of the exclusive $\Delta_{1232}$
cross section the quality of the data and thus also the quality
of the resulting fit is very good. For higher resonance excitations
this is unfortunately no longer the case and additional measurements 
are needed to clarify the situation.
One has to keep in mind, however, that the experimental extraction
of exclusive resonance production cross sections is only possible
via two- or three-particle correlations (e.g. a pion-nucleon correlation
in the case of the $\Delta$) which introduces large systematic errors,
especially for broad resonances.

\begin{figure}
\centerline{\psfig{figure=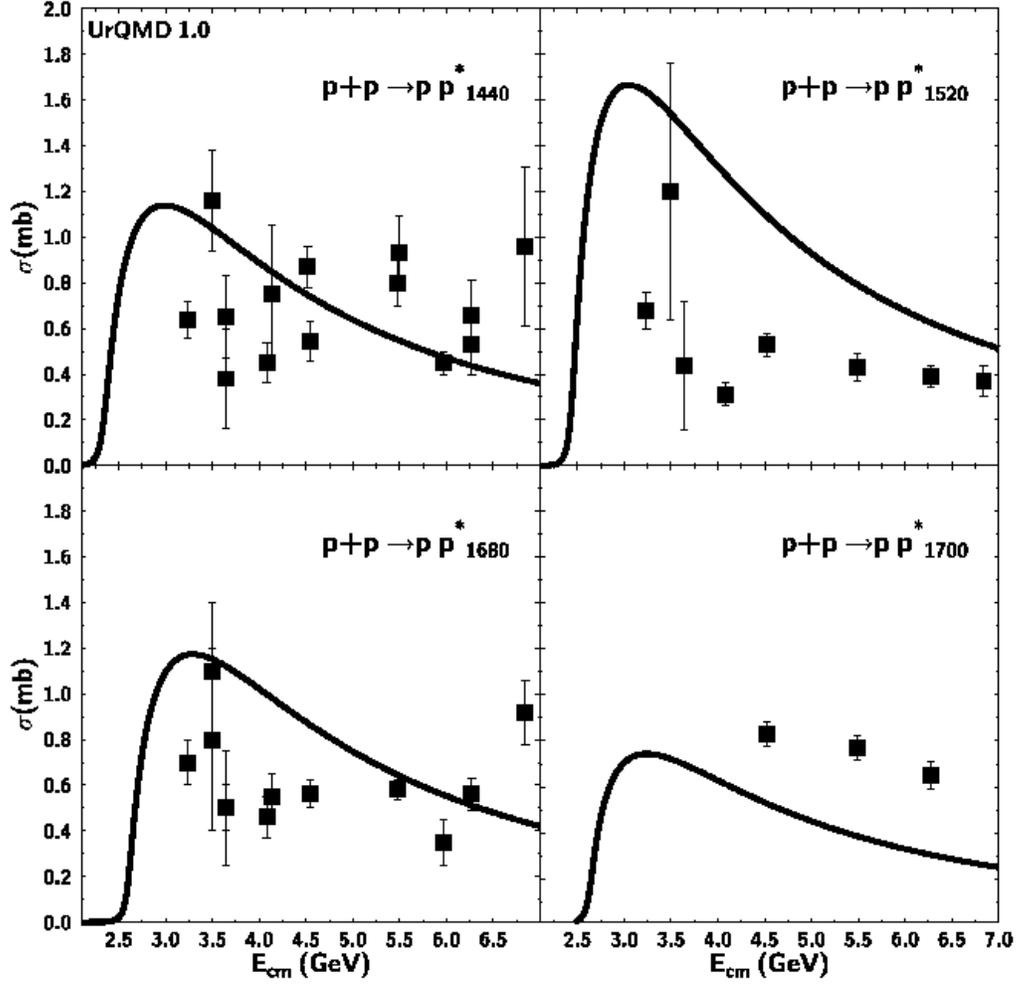,width=15cm}}
\caption{\label{pp-ppstar}
UrQMD parameterization for exclusive $p p^*$ cross sections.
Only one parameter was used to describe all  available 
cross section data \protect\cite{flaminio}. 
}
\end{figure}

In figure~\ref{pp-ppstar} the UrQMD cross sections for the processes
$p p \to p p^*_{1440}$, $p p \to p p^*_{1520}$
$p p \to p p^*_{1680}$ and $p p \to p p^*_{1700}$
are compared to data \cite{flaminio}.
One single parameter has been used to fit all four cross sections.
The data situation is not as good as in the case of the
$\Delta_{1232}$ resonance, some ambiguities are visible which results in
the quality of the fit being not as good as in the previous case. 
The parameters for the other classes are fitted in the same fashion.

\begin{figure}
\centerline{\psfig{figure=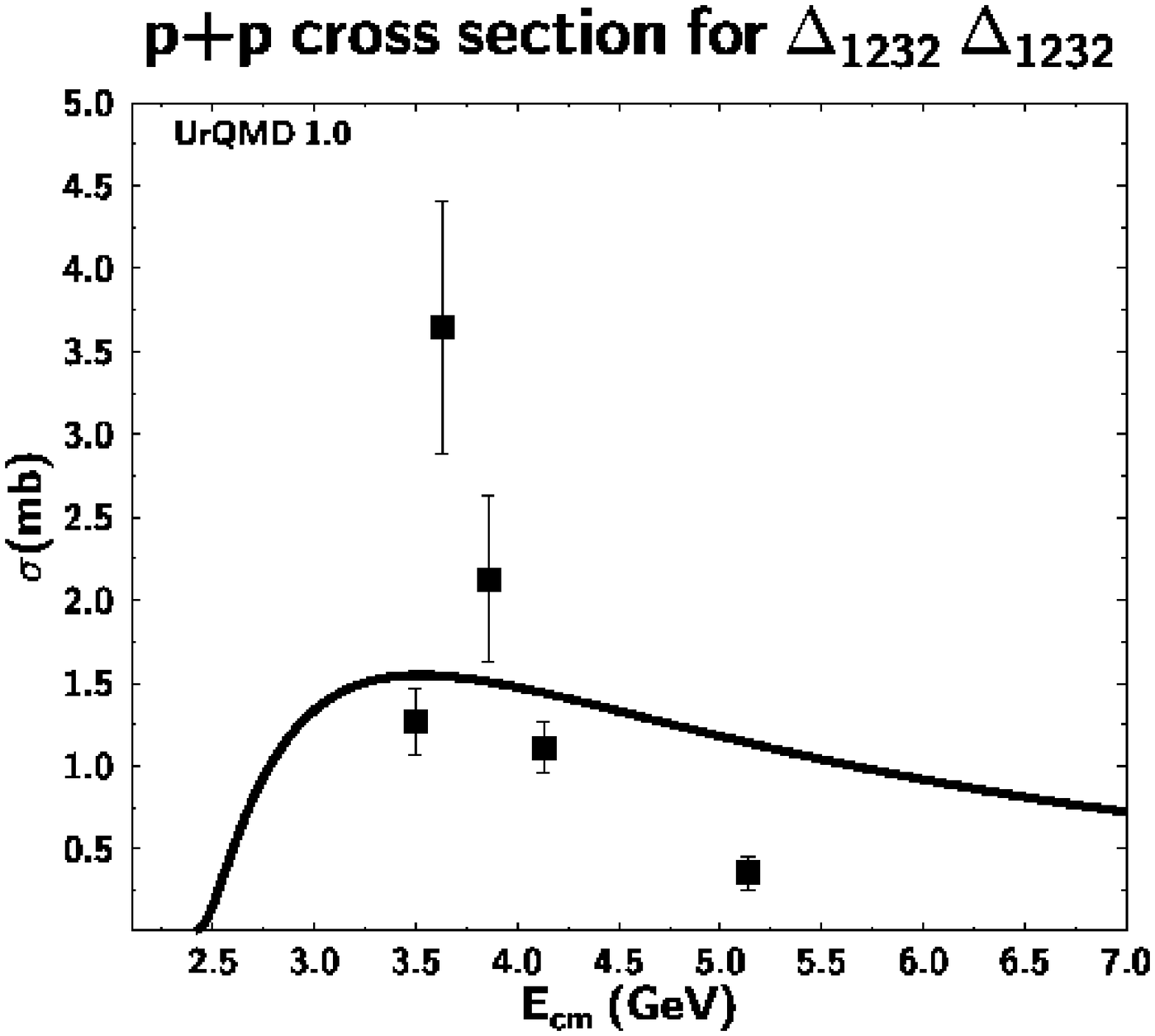,width=15cm}}
\caption{\label{pp-2d1232}
UrQMD fit for the exclusive $\Delta_{1232} \Delta_{1232}$ 
production in proton-proton
reactions  compared to data
\protect\cite{flaminio}. }
\end{figure}

The cross section for exclusive $\Delta_{1232} \Delta_{1232}$
can be seen in figure~\ref{pp-2d1232}. The data points 
\cite{flaminio} hint at a 
resonance like structure which cannot be reproduced with
the UrQMD ansatz for resonance-excitation cross sections.
However, the data deviates considerably from other cross sections for
resonance excitation (e.g. $N N \to N^* \Delta^*$). Considering
hadron-universality and the similarities in all other resonance excitation
cross sections this casts a certain doubt on the accuracy of the
measurement of the resonance-like peak.

\begin{figure}
\centerline{\psfig{figure=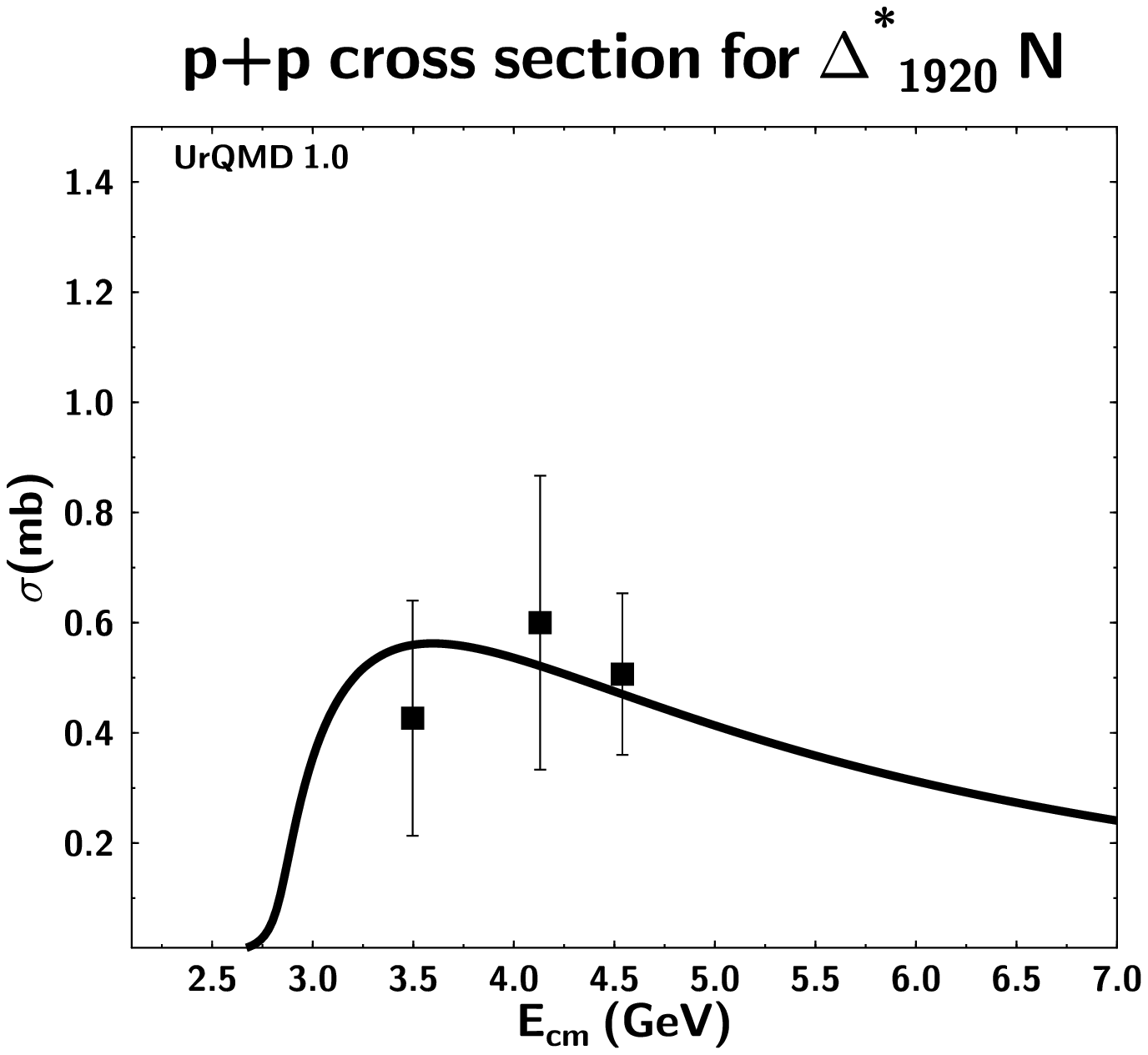,width=15cm}}
\caption{\label{pp-nd1920}
UrQMD fit for the exclusive $\Delta_{1920} N$ 
production in proton-proton
reactions  compared to data
\protect\cite{flaminio}. }
\end{figure}

\begin{figure}
\centerline{\psfig{figure=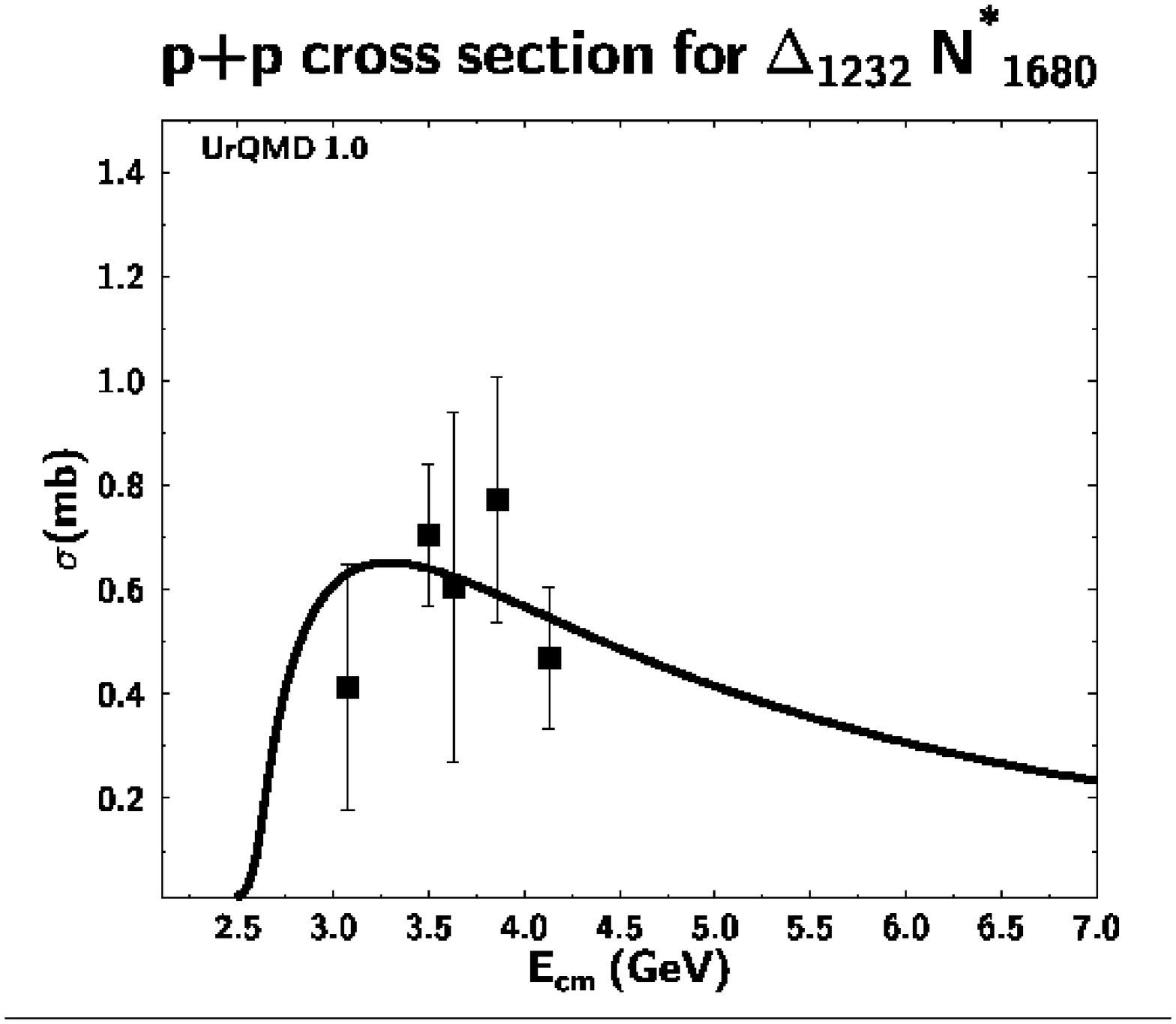,width=15cm}}
\caption{\label{pp-d1232n1680}
UrQMD fit for the exclusive $\Delta_{1232} N^*_{1680}$ 
production in proton-proton
reactions  compared to data
\protect\cite{flaminio}. The matrix element for
all $\Delta_{1232} N^*_X$ reactions is extracted from this fit.}
\end{figure}

\begin{figure}
\centerline{\psfig{figure=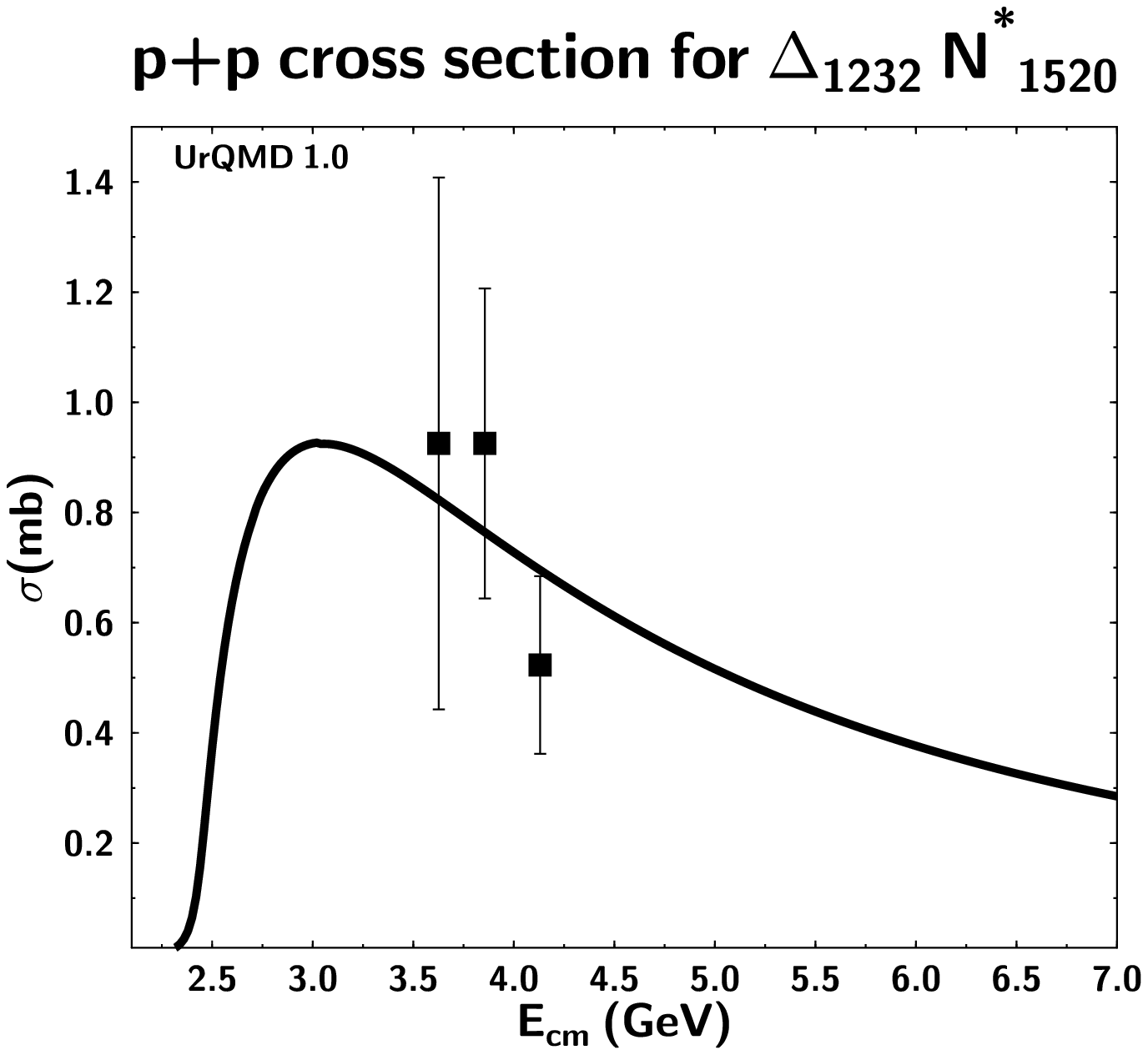,width=15cm}}
\caption{\label{pp-d1232n1520}
Comparison between the UrQMD parametrization for the 
exclisve $\Delta_{1232} N^*_{1520}$ production in proton-proton
reactions  compared to data
\protect\cite{flaminio}. The matrix element has been extracted
from a fit to the exclusive $\Delta_{1232} N^*_{1680}$ production.} 
\end{figure}

Figure~\ref{pp-nd1920} shows the UrQMD fit for the exclusive
$\Delta^*_{1920} N$ production. The same matrix element is
used for the entire class of $N N \to \Delta^* N$ reactions.
In the case of exclusive $\Delta^* N^*$ production 
the matrix element has been fitted to the
$\Delta^*_{1232} N^*_{1680}$ exit channel (see figure~\ref{pp-d1232n1680}).
The extrapolation to the case of  $\Delta^*_{1232} N^*_{1520}$
production can be seen in figure~\ref{pp-d1232n1520}.
For the exclusive $p p \to \Delta \Delta^*$ reaction class the data
situation is unsatisfactory, therefore we used the same matrix-element
as in the $p p \to N \Delta^*$ case.


The decay of the resonances proceeds according to the
branching ratios compiled by the Particle Data Group \cite{pdg96}.
The resonance decay products have isotropical distributions in the rest
frame of the
resonance. If the resonance decays into $N > 2$ particles, then
the corresponding $N-$body phase space is used to calculate their $N$
momenta stochastically.
It is necessary to note that a consistent description of angular
momentum distributions points to a rather intricate problem of transport
theory itself: If one considers the whole
scattering interaction to be described by one single quantum mechanical
process
there are correlations between the final and the initial stage. For
instance the angular distribution  of the final particles with respect
to the axes of the incoming momenta  in the CMS system. 
However, a fitting of the angular distributions to experimental data 
may  conflict with the basic assumption of transport
theories that the multiple scattering processes can be considered to be
of Markovian type, i.e. after each scattering process or resonance
formation the outgoing particles completely forget about their entrance
channels. In the case of a spin 0 resonance there is no
preferred direction for the emission of the final particles, while for
spin 1 (and other) the different magnetic quantum numbers are
statistically occupied, so that also in these cases there is no
preferred angle of emission\footnote{For a detailed discussion of the
influence of non-markovian processes in the transport theory of heavy
ion collisions we refer the reader to Refs. \cite{dan84,gr92}.}.

All produced particles are able to rescatter within the nuclear
medium, therefore the excitation of resonances by the
annihilation of mesons on baryons included as depicted in Fig.~\ref{fig5} for
the reaction $\pi^+ + p\rightarrow \Delta^{++(*)}$.

Also the $\pi^- + p$ channel (Fig.~\ref{fig6}) shows a rich structure of
baryon resonances. 
The total meson-baryon cross section is given by formula~\ref{mbbreitwig}.
There, the total and partial decay widths also define the inverse 
reaction, i.e. the different decay-channels of the respective resonance.
Thus, the principle of detailed balance is applied.
Based on this principle 
we calculate all resonance formation cross sections from the
measured decay properties of the possible resonances up to
c.m. energies of 2.25~GeV/$c^2$ for baryon resonance and 1.7~GeV/$c^2$
in the case of MM and MB reactions. Above this energy
collisions are modeled by the formation of $s$-channel string or, at
higher energies (beginning at $\sqrt s =3$~GeV), by one/two $t$-channel
strings. In the strangeness channel
elastic collisions are possible for those meson-baryon combinations
which are not able to form a resonance, while the creation of
$t$-channel strings is always possible at sufficiently large
energies (c.f. Fig.~\ref{fig7} for the formation of hyperon resonances
and Fig.~\ref{fig8} for the non-resonant channel). At high collision
energies both cross section become equal due to quark counting rules.

In more general terms, the principle of detailed balance 
can be derived by assuming time-reversal
invariance of the interaction Hamiltonian and
can be formulated in the following way:
\be
\label{detbal}
\sigma(y\rightarrow x)\, p_y^2 \, g_y =
\sigma(x\rightarrow y)\, p_x^2 \, g_x\quad,
\ee
with $\vec{p}$ denoting the c.m.-momenta of the particles and $g$ being
the spin-isospin degeneracy factors.
Thus, if the cross section of the reaction
$x\rightarrow y$ is known, the back reaction $y\rightarrow x$ can
be easily obtained. This principle is in UrQMD widely applied for
the calculation of baryon-resonance absorption cross sections,
such as $\Delta(1232) + N \to N + N$. For a detailed discussion
of the application of the principle of detailed balance to resonance
absorption and $\sqrt{s}$-dependent decay widths we refer to~\cite{urqmd}.

As was mentioned above, not only baryon-baryon and meson-baryon
collisions have to be included in the proposed scheme. At high
energies and in massive AA systems meson-meson
collisions may dominate the multiple production of
secondaries. Unfortunately, there are only few channels for which the
experimental information exists, like the process of $\pi^+\pi^-$
scattering (Fig.~\ref{fig9}), which is fairly described by the
UrQMD model.

\subsection{Strings}

Both, for the high energy regime and for baryon$-$antibaryon
annihilation we apply a string model (similar to the LUND model \cite{boa})
to describe the inelastic
reactions. The constituents, quark and diquark (or anti-quarks),
of the incoming hadron also define the predominant emission patterns of
the events. The amount of stopping in nucleus-nucleus is strongly
correlated to the detailed dynamics of the diquark in the hadronic
medium \cite{urqmd}. 
Recently different additional mechanisms of baryon number transport in
nucleus-nucleus reactions have been investigated: Baryonic junctions
as suggested by \cite{kopel} which yield an enhanced hyperon and proton
production cross section at central rapidities \cite{vance}; 
di-quark breaking due to interaction with the hadronic medium 
as predicted by \cite{acap}. The di-quark breaking component is also
taken into account in the UrQMD model (di-quark breaking probability$ =
10\%$). However, the above mechanisms are of minor importance in the
UrQMD approach since rescattering of the leading di-quark with hadrons
is explicitly taken into account.

Since gluons are massless particles with spin $J = 1$,
the static strong interaction between quarks at small distances $(r \ll 1/
\Lambda_{\rm QCD})$ may be described by a potential $V_0 \propto
- \alpha_S/r$. At sufficiently large distances the color field between
two quarks or anti-quarks transforms into the color string.
Denoting the string tension as $\kappa$ one defines the linear string
potential
\begin{equation}
V_1=\kappa\,|z_1-z_2|\quad,
\end{equation}
between the quarks/diquarks located at $z_1$ and $z_2$, respectively.
This form of the potential is chosen from heuristic considerations,
based on the quark confinement hypothesis,
and is supported by lattice QCD calculations \cite{lattice}.

The transverse directions have not to be taken into account, since
they are negligibly small compared to the longitudinal excitation of
the hadron string. Hence, we get the dynamics of the quark system
(with quark momenta $p_1$ and $p_2$) from the Hamiltonian $H$
\begin{equation}
H=|p_1|+|p_2|+\kappa|z_1-z_2|\quad ,
\end{equation}
which leads to the following equations of motion for the massless
endpoints of the string:
\begin{eqnarray}
\frac {{\rm d} p_i}{{\rm d}t} & = &
           -\frac {\partial H}{\partial z_i}  =
        - {\rm sign}(z_i - z_i')\,\kappa \quad ,\\
\frac {{\rm d} z_i}{{\rm d} t} & = &
           +\frac {\partial H}{\partial p_i}  =
          {\rm sign}(p_i) \quad.
\end{eqnarray}
A change in momentum is directly related to the sign of $(z_i-z_i')$,
while the direction of propagation is defined by the sign of the
momentum $p_i$ of the quark. This results in a typical ''yo-yo''
type evolution of the quark system.

If the momentum transfer is large enough, the excitation of the
string may exceed some critical limit. After that it will be
energetically favorable to break the string into pieces by producing
$q \bar q$-pairs from the vacuum. Each of the produced
$q \bar q$-pairs will have small relative momenta in their rest
frame. Owing to the fact that the color string is uniformly stretched,
the hadrons produced as a result of the string fragmentation will be
uniformly distributed within the kinematically allowed interval
between $y_{min} = 0$ and $y_{max} = \ln{(s/m_T^2)}$.

The probability of the pair
production process has been calculated by Schwinger for the case of
an infinite homogeneous electrical field. His result can be used to
motivate the decay of QCD color field between the quarks.
Note that QCD is a non-abelian theory, therefore the
color field need not be homogeneous and it is definitely not infinite. This
leads to the modifications of string decays which will be discussed later.

\begin{figure}
\centerline{\psfig{figure=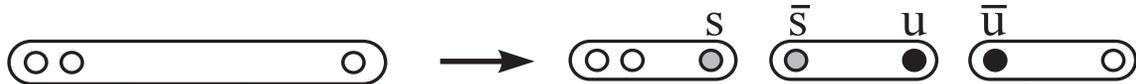,width=15cm}}
\caption{Scheme of a decaying string. $s\bar s$ and $u\bar u$ pairs
are created in the color field resulting in a hyperon, a kaon and a pion.
}
\label{fig20}
\end{figure}

The probability $|M|^2$ for the creation of a quark-antiquark pair
with mass $m$ in a color field with a string tension $\kappa$ is:
\begin{equation}
|M|^2= {\rm constant} \times \exp{\left( -\frac{\pi m^2}{\kappa}
\right) } \quad,
\end{equation}
where a typical value for $\kappa$ is 1~GeV/fm. 
This relation is
motivated by Schwinger's QED-based result for particle-antiparticle creation
in a strong electric field (see the discussion of formula~\ref{schw} 
in section~\ref{finite}).
The relative production probabilities of the different quark flavors are
adjusted to $e^+e^-$-data:
\begin{equation}
u:d:s:\mbox{diquark} = 1:1:0.35:0.1\quad.
\end{equation}
The production of charmed (and heavier) quarks is strongly suppressed in
the string picture, hence they are exclusively produced in hard QCD
processes.
The elementary diquark is introduced to allow for baryon-antibaryon
production in the string. A schematic view of the decaying string is
shown in Fig.~\ref{fig20} $-$ a non-strange baryon decays into a
hyperon, a kaon and a pion.

To decide which type of hadron is produced from the quark configuration
that is created in the color field we choose in the case of a
produced

({\bf i}) {\it baryon\/} $-$ the octet and decuplet with equal
probabilities;

({\bf ii}) {\it meson\/} $-$ the meson nonet with a probability
proportional to the spin degeneracy and inverse mean mass $m$,
\begin{equation}
P_{\rm multiplet} \propto \frac{2S+1}{\langle m\rangle_{\rm multiplet}}\quad.
\end{equation}
The singlet states are mixtures of $u\bar u$, $d\bar d$ and $s\bar
s$. They are projected onto SU$(3)$ hadrons with the flavor mixing
angles from the quadratic Gell-Mann-Okubo mass formula \cite{gellm}. 
For the scalar mesons this formula is not
applicable, here an ideal mixing angle (${\rm tan}(\theta)=1/\sqrt{2}$)
is assumed (The mixing angles are depicted in Table \ref{mixang}). 

The Field$-$Feynman fragmentation mechanism
\cite{feynman}, which allows the independent string decay from both
ends of the string is used in the UrQMD model.  The string break-up is treated
iteratively: String $\rightarrow$ hadron + smaller string.
The conservation laws are fulfilled. The diquark is permitted to
convert into mesons via the breaking of the diquark link, thus
transporting the baryon number into central rapidities.

On both sides of the fragmenting string the new particles are formed
randomly. If
a resonance is produced, its mass is determined by a Breit-Wigner mass
distribution. The transverse momentum is assigned to this particle
according to a thermal momentum distribution, resulting in a temperature
of 170~MeV. After that the fragmentation
function determines the fraction of the longitudinal momentum of the string
transferred to the hadron. This procedure can be described in a
covariant manner by the light cone variables defined as:
\begin{equation}
z^{\pm } = t \pm z \quad {\rm and} \quad p^{\pm } = E \pm  p \quad .
\end{equation}
The light cone momentum $p^{\pm }_{\rm hadron}$ given to the newly
produced hadron is:
\begin{equation}
p^{\pm }_{\rm hadron}=z^{\pm}_{\rm fraction}\, p^{\pm }_{\rm total}
\end{equation}
The fragmentation of a baryonic string reads:
\begin{equation}
\label{fragbarstr}
    p^- \underbrace{(qq\, q \bar q \,  q )}_{\mbox{String}}
  =   z^{-}_{\rm fraction} p^-\underbrace{( qqq  )}_{\mbox{Baryon}}
                             +\,  (p^--z^{-}_{\rm fraction}p^-)
       \underbrace{\bar q q}_{\mbox{String}} \qquad.
\end{equation}

\begin{figure}
\centerline{\psfig{figure=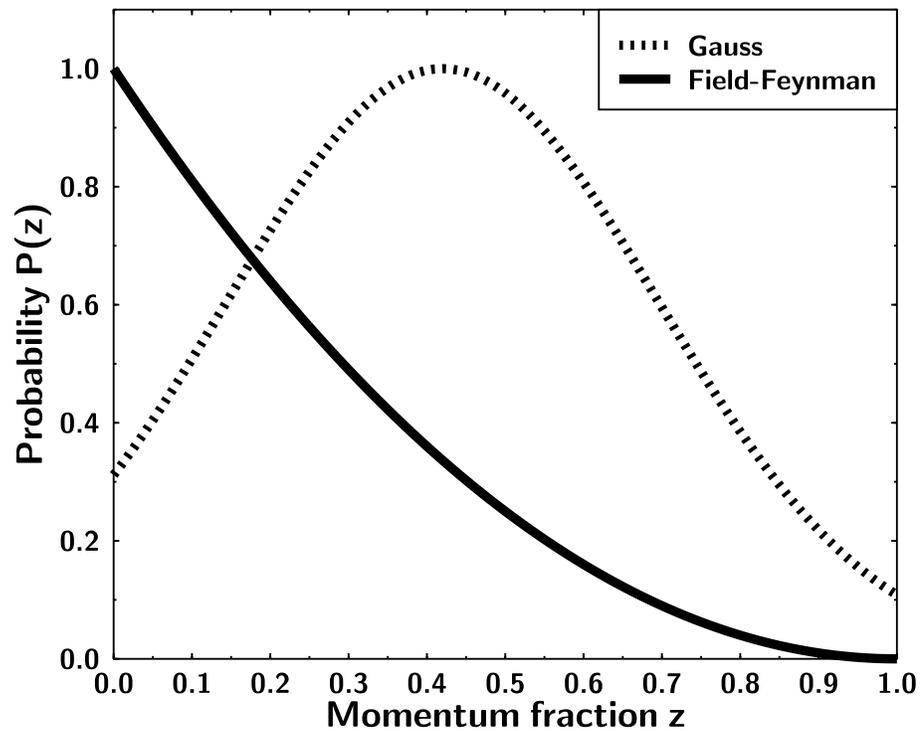,width=15cm}}
\caption{The Field-Feynman fragmentation function (solid line) is used for newly
produced particles. A Gaussian fragmentation function (dashed line)
is used for leading baryons.
}
\label{fig21}
\end{figure}

\begin{figure}
\centerline{\psfig{figure=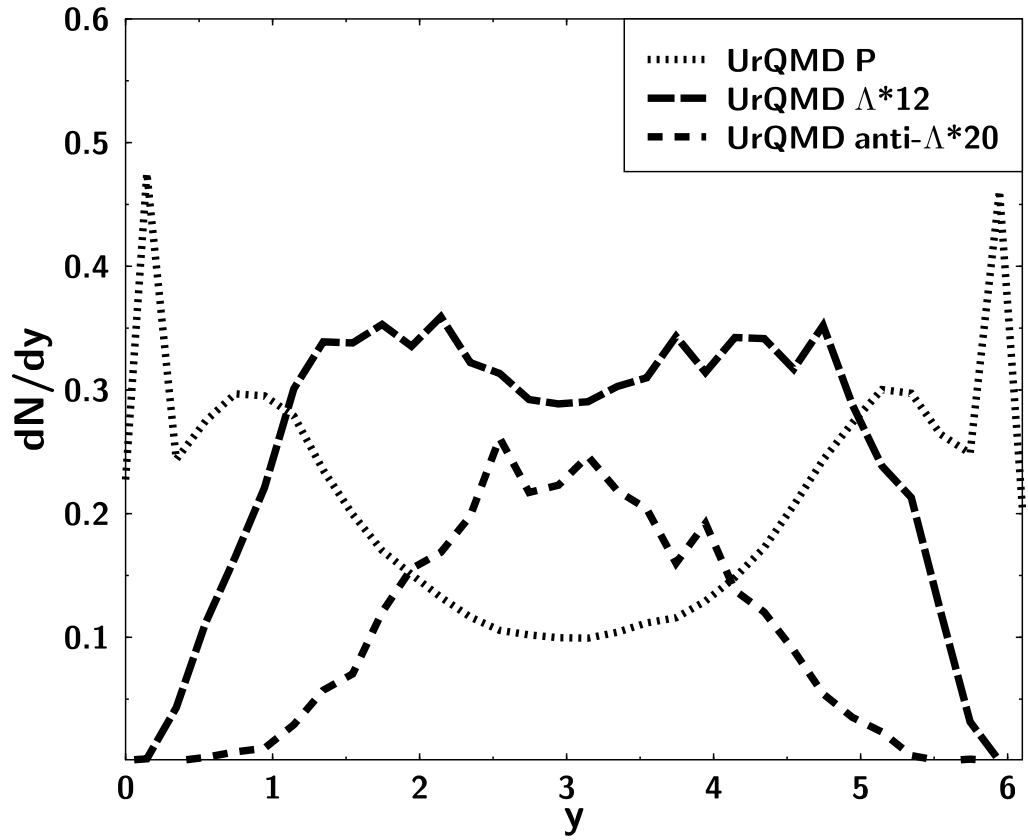,width=15cm}}
\caption{Rapidity spectrum of protons (dotted curve), $\Lambda$'s
(dashed) and $\bar \Lambda$'s (dash-dotted) for $pp$ collisions at
205 GeV/$c$. Data are taken from \protect\cite{kafka}. 
}
\label{fig22}
\end{figure}

\begin{figure}
\centerline{\psfig{figure=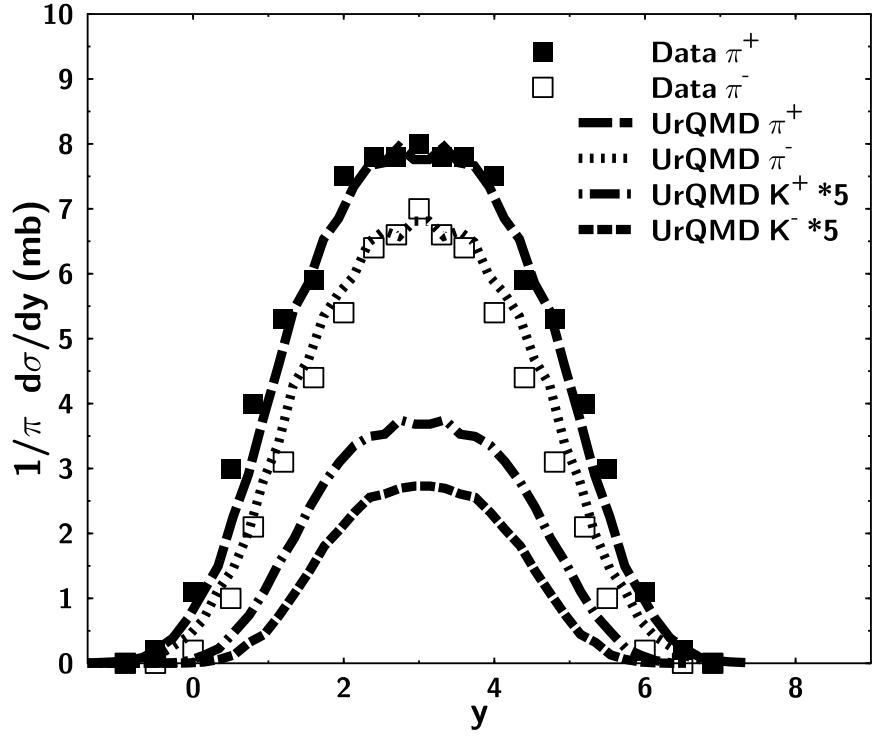,width=15cm}}
\caption{Rapidity distribution of $\pi^+$ and $\pi^-$ in $pp$
collisions at 205 GeV/$c$. Data are taken from \protect\cite{kafka}.}
\label{fig23}
\end{figure}

The main input is the fragmentation function which yields the
probability distribution $p(z^{\pm}_{\rm fraction}, m_t)$.
This function regulates the fraction of energy and
momentum given to the produced hadron in the stochastic
fragmentation of the color string.
For newly produced particles the Field-Feynman function
\cite{feynman}:
\begin{equation}
p(z^{\pm}_{\rm fraction})={\rm constant} \times (1-z^{\pm}_{\rm fraction})^2,
\end{equation}
is used.
$P(z)$ drops rapidly with increasing $z$ (Fig.~\ref{fig21}).
Therefore, the longitudinal momenta of e.g. produced antibaryons
(Fig.~\ref{fig22})
and  pions (Fig.~\ref{fig23}) are small (they stick to central rapidities),
in line with the experimental data.
The rapidity spectra of these particles have a characteristic
Gaussian-like shape, in contrast to the baryon spectra in pp, as it is
clearly seen in Figure \ref{fig22}.

\begin{figure}
\centerline{\psfig{figure=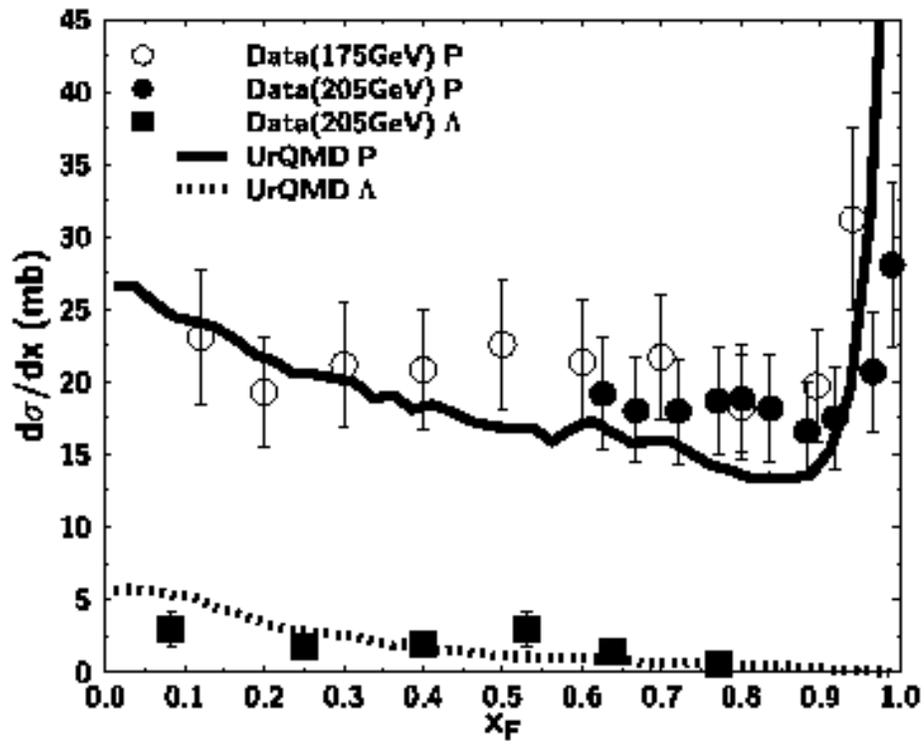,width=15cm}}
\caption{${\rm d} \sigma / {\rm d} x_F$ distribution of protons and
$\Lambda$'s in $pp$ collisions at 205 GeV/$c$. 
Data are taken from \protect\cite{kafka}.}
\label{fig24}
\end{figure}

The proton is on  average less stopped, since it is build up from the leading
diquark in the string (leading particle effect). Fig.~\ref{fig24} compares
the $x_F$
distribution of protons and $\Lambda$'s for the Feynman scaling variable
$x_F=2 p_{||}/\sqrt{s}$ measured in pp reactions at
205~GeV/$c$. The data on leading baryons can only be reproduced when a
modified
fragmentation function is used for the leading baryons (cf.
Fig.~\ref{fig21}, dashed curve). This leading baryon fragmentation function
is of
Gaussian form:
\begin{equation}
p(z^{\pm}_{\rm fraction})= {\rm constant} \times \exp{\left[-\frac
{(z^{\pm}_{\rm fraction}-b)^2}{2a^2} \right] }\quad,
\end{equation}
with parameters $a = 0.275$ and $b = 0.42$.

It is obvious that modeling the momentum loss in elementary
collisions has a strong influence on the rapidity spectra of particles
produced in heavy-ion collisions. This can be seen if one compares
the spectra of p, $\Lambda$'s (created from the leading baryon)
and $\bar \Lambda$'s or mesons (created from newly produced quarks)
in lead-lead collisions at the SPS energies \cite{urqmd}.

\subsubsection{Finite Size Effects}
\label{finite}

Let us now discuss finite size effects in the
process of string fragmentation. The string is essentially a color field which
connects two color charges, the $[\, 3\,]$ quark and the $[\,\overline 3\,]$
di-quark or
anti-quark, at the ends of the string. Quantum electrodynamics (QED) predicts
spontaneous particle$-$antiparticle creation in strong electric
fields \cite{hei36,sch51}. This effect should hold for particle
creation in a strong color fields. Schwinger's QED result
\be
W_\infty=\frac{(eE)^2}{4\pi^3}\sum^\infty_{N=1}
\frac{1}{N^2} \exp{\left(-N\pi\frac{m^2}{eE}\right)}
\label{schw}
\ee
is often adopted
to the case of color fields by equating $|eE|$ with the string tension
$\kappa$.

However, for strings, several important assumptions, which lead to
Schwinger's result are not fulfilled: Firstly, the color field is not
infinitely extended. It is bound radially by the interaction length of
the gluons and longitudinally roughly by the $[\, 3\,]$ and $[\,\overline 3\,]$
endpoints. Secondly,
the two endpoints of the string move (with close to the speed of light) in
opposite directions. Finally, the assumption of a constant field
strength $\kappa$ seems to be fulfilled along the string, but may
become invalid in hot and dense matter \cite{ropes}.

The influence of the finite radial size
of the string can be studied by solving the Dirac equation for the
(newly produced) pairs in a finite volume. The field is restricted to a
cylindrical volume of length $L$ and and area $\pi R^2$, where $R$ is
the cylinder radius. Along the cylinder axis a
homogeneous color field is assumed. The boundary condition on the
surface of the cylinder leads to enclosure of the color charges.
Taking $L\rightarrow\infty$ neglects the longitudinal direction.

The constraint on the cylinder surface is given by linear MIT-boundary
conditions \cite{cho73}, thus we have to solve the equations:
\begin{eqnarray}
(\gamma_\mu p^\mu -e\gamma_\mu A^\mu -m)\psi(x) &=& 0 \quad, \label{dirac1}\\
(i n^\mu \gamma_\mu -1)                 \psi(x) &=& 0 \quad
(x\in\partial V)\quad.\label{dirac2}
\end{eqnarray}
The MIT model allows for an analytic solution\cite{tomplb}.
The occupation numbers in the limits $t\rightarrow
\pm\infty$ yield the pair creation probability as \cite{tomplb}
\be
W_R=\frac{(eE)^2}{4\pi^3}\sum^\infty_{N=1}
\frac{1}{N^2}\exp{\left(-N\pi\frac{m^2}{eE}\right)}
\times
\left\{
\frac{2\pi N}{eER^2}\sum_{n,\mu}\exp{\left[ -N\pi\frac{(k_{n\mu} R)^2}
{eER^2} \right] } \right\}\quad .
\ee
Thus the string radius is directly related to the pair creation rate
$W_R$, with $k_{n\mu}$ being the n$^{\rm th}$ momentum eigenvector in the 
solution of Eqs. (\ref{dirac1} and \ref{dirac2}) for a given projection $\mu$ of the
corresponding Bessel functions.

\begin{figure}
\centerline{\psfig{figure=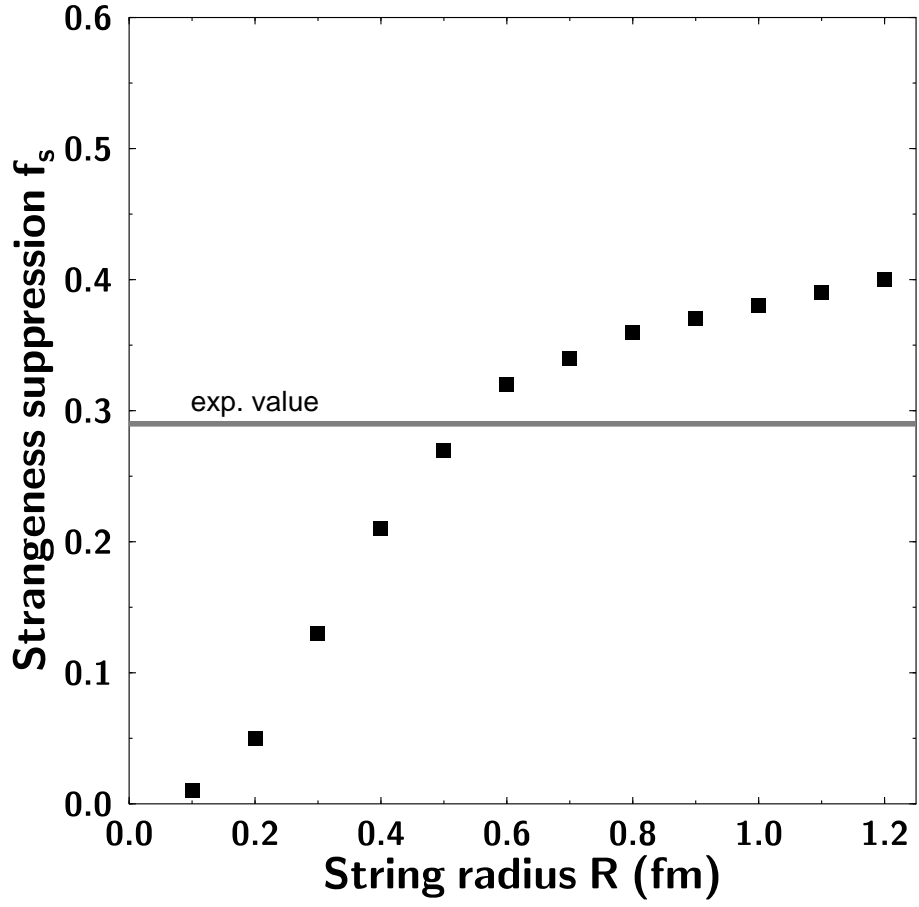,width=15cm}}
\caption{Strangeness suppression $f_s$ due to a finite transverse string
radius $R$ \protect\cite{tomplb}.}
\label{fig26}
\end{figure}

The expression in curly brackets gives the deviation from the Schwinger
formula. For $R\rightarrow\infty$ this second factor
vanishes and one is back to the infinite case. This result allows
to calculate the strangeness suppression
$f_s(R) = W_R(s)/W_R(u)$ for different string radii as shown in
Fig.~\ref{fig26}.

\begin{figure}
\centerline{\psfig{figure=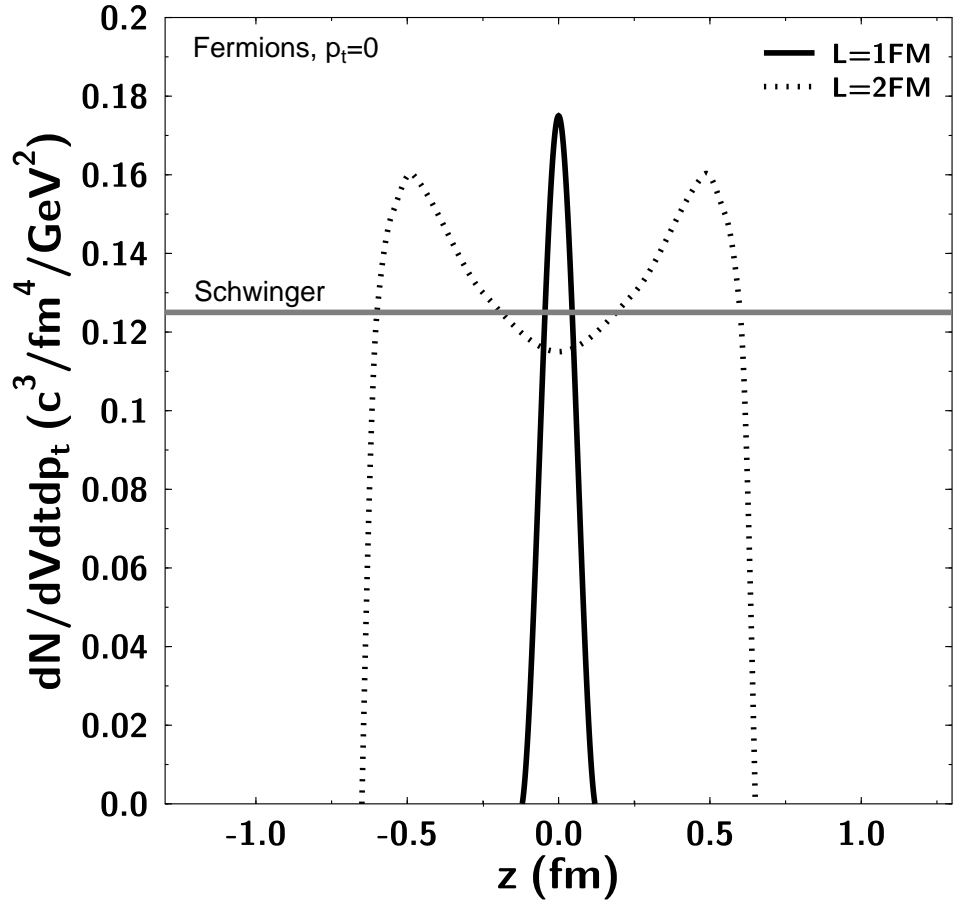,width=15cm}}
\caption{Pair production rate in a finite color field as a function of
the longitudinal field extension $z$ \protect\cite{wang}.}
\label{fig27}
\end{figure}

The overall pair creation rate has been calculated as a function of the
longitudinal size of the color field for the corresponding Dirac
equation \cite{wang} in terms of confluent
hypergeometric functions. The resulting particle production rate per volume
d$V$ and time interval d$t$,
${\rm d}N/({\rm d}V{\rm d}t{\rm d}p_T)$ at $p_T=0$ is shown in
Fig.~\ref{fig27} as a function of $z$, the distance from the string
center, for different string lengths $L$. A clear depletion of
particle production near the endpoints of the string endings  is
visible (Eq.\ref{schw}). For short strings the
particle production gets enhanced toward the string center.

The above discussed finite size effects have a strong influence on the results,
especially on the production of heavy quarks and particles like anti-baryons,
-hyperons, etc. Up to now these corrections are included only
in the non-strange antibaryon sector. This effect is very important
especially in the case of the $\Lambda$-particle, since it is mostly
created at the string ends in the fragmentation of a leading proton.
Indeed, it has been reported that most of the transport models which
use a string fragmentation scheme based on the Schwinger formalism
tend to overestimate the $\Lambda$'s \cite{pop}. Therefore, this finite
size effects shall be included
for strange baryons, in order to correct the overestimated number of
$\Lambda$'s and $\bar{\Lambda}$'s.

\section{The Generation of Transverse Momentum}
\label{sec5}

In hot and dense nuclear matter, most hadrons suffer interactions from
the many surrounding particles. As a result, the effective mass may change
with density. Many dynamical properties of hadrons are modified
in the
medium. In-medium two-body scattering cross-sections may differ
from the free space values. Those effects can be studied in the
framework of relativistic transport theory, i.e. the relativistic
Boltzmann-Uehling-Uhlenbeck (RBUU) equation. This type of transport
equation has been used extensively to the study of relativistic
heavy-ion collisions and turned out to be very successful. The UrQMD
approach uses an analytical expression for the differential
cross-section of in-medium $NN$ elastic scattering derived from the
collision term of the RBUU equation \cite{Mao} to determine the
scattering angles between the outgoing particles in elementary
hadron-hadron collisions. It is assumed that the angular
distributions for all relevant two-body processes are similar modified
in an analogous manner. They are approximated by the differential
in-medium $NN$ elastic scattering cross-section:
\begin{equation}
  \sigma_{NN \rightarrow NN}(s,t) = \frac{1}{(2 \pi)^{2} s}
  \lbrack D(s,t)
  + E(s,t) + (s, t \longleftrightarrow u) \rbrack,
\label{angeq}  \end{equation}
with the direct term
 \begin{eqnarray}
  D(s,t)&=& \frac{({\rm g}_{NN}^{\sigma})^{4}}{2(t-m_{\sigma}^{2})^{2}}
 (t- 4 m^{*2})^{2} + \frac{({\rm g}_{NN}^{\omega})^{4}}{(t-m_{\omega}^{2})^{2}}
 (2 s^{2} + 2st +t^{2} -8m^{*2}s +8m^{*4}) \nonumber \\
  &&  + \frac{24({\rm g}_{NN}^{\pi})^{4}}{(t- m_{\pi}^{2})^{2}} m^{*4}t^{2}
 - \frac{4({\rm g}_{NN}^{\sigma}{\rm g}_{NN}^{\omega})^{2}}
{(t - m_{\sigma}^{2})(t-m_{\omega}^{2})} (2s + t -4m^{*2})m^{*2},
\end{eqnarray}
and the exchange term
\begin{eqnarray}
  E(s,t)&=& -\frac{({\rm g}_{NN}^{\sigma})^{4}}{8(t-m_{\sigma}^{2})
 (u-m_{\sigma}^{2})} \lbrack t (t+s) + 4m^{*2}(s-t) \rbrack
  + \frac{({\rm g}_{NN}^{\omega})^{4}}{2(t-m_{\omega}^{2})(u-m_{\omega}^{2})}
 (s- 2m^{*2}) \nonumber \\ && \times (s-6m^{*2})
    - \frac{6({\rm g}_{NN}^{\pi})^{4}}{(t- m_{\pi}^{2})(u-m_{\pi}^{2})}
  (4m^{*2}-s -t ) m^{*4}t \nonumber \\
  && + ({\rm g}_{NN}^{\sigma}{\rm g}_{NN}^{\pi})^{2}
 \lbrack \frac{3m^{*2}(4m^{*2}-s-t)(4m^{*2}-t)}{2(t-m_{\sigma}^{2})
(u-m_{\pi}^{2})} + \frac{3t(t+s)m^{*2}}{2 (t-m_{\pi}
^{2})(u-m_{\sigma}^{2})}  \rbrack \nonumber \\
&& + ({\rm g}_{NN}^{\sigma}{\rm g}_{NN}^{\omega})^{2}
 \lbrack \frac{t^{2} - 4m^{*2}s -10m^{*2}t +24m^{*4}}{4(t-m_{\sigma}^{2})
(u-m_{\omega}^{2})} + \frac{(t+s)^{2} - 2m^{*2}s + 2m^{*2}t }{4 (t-m_{\omega}
^{2})(u-m_{\sigma}^{2})}  \rbrack \nonumber \\
  && + ({\rm g}_{NN}^{\omega}{\rm g}_{NN}^{\pi})^{2}
 \lbrack \frac{3m^{*2}(t+s-4m^{*2})(t+s-2m^{*2})}{(t-m_{\omega}^{2})
(u-m_{\pi}^{2})} + \frac{3m^{*2}(t^{2}-2m^{*2}t)}{ (t-m_{\pi}
^{2})(u-m_{\omega}^{2})}  \rbrack,
  \end{eqnarray}
The (pseudo-)scalar and vector coupling constants  are ${\rm
g}_{NN}^{\pi}=1.434$, ${\rm g}_{NN}^{\sigma}=6.9$,
and ${\rm g}_{NN}^{\omega}=7.54$ and
$m^*$ is the in-medium mass,
$s, t, u$ are the Mandelstam variables.
The
in-medium single-particle energy is given by
\begin{equation}
E^{*}(p)=\sqrt{{\bf p}^{*2}+m^{*2}} \; .
\end{equation}
The formula for the differential cross section of in-medium $NN$
elastic scattering can be used for elementary hadron-hadron
collision if it is scaled by
\begin{equation}
s \rightarrow s-(m_1^{*}+m_2^{*})^2+4m^{*^2} \; ,
\end{equation}
where $m_1^{*}$ and $m_2^{*}$ denote the effective masses of the
incoming hadrons. Furthermore, we take into account finite size effects
of the hadrons and part of the short range
correlation by introducing a phenomenological form factor at each
vertex. For the baryon-baryon-meson vertex the common
form
\begin{equation}
F_{BBM} = \frac{\Lambda_M^2}{\Lambda_M^2-t} \; .
\end{equation}
is used, where $\Lambda_M$ is the cut-off mass for meson $M$.\\

\begin{figure}
\centerline{\psfig{figure=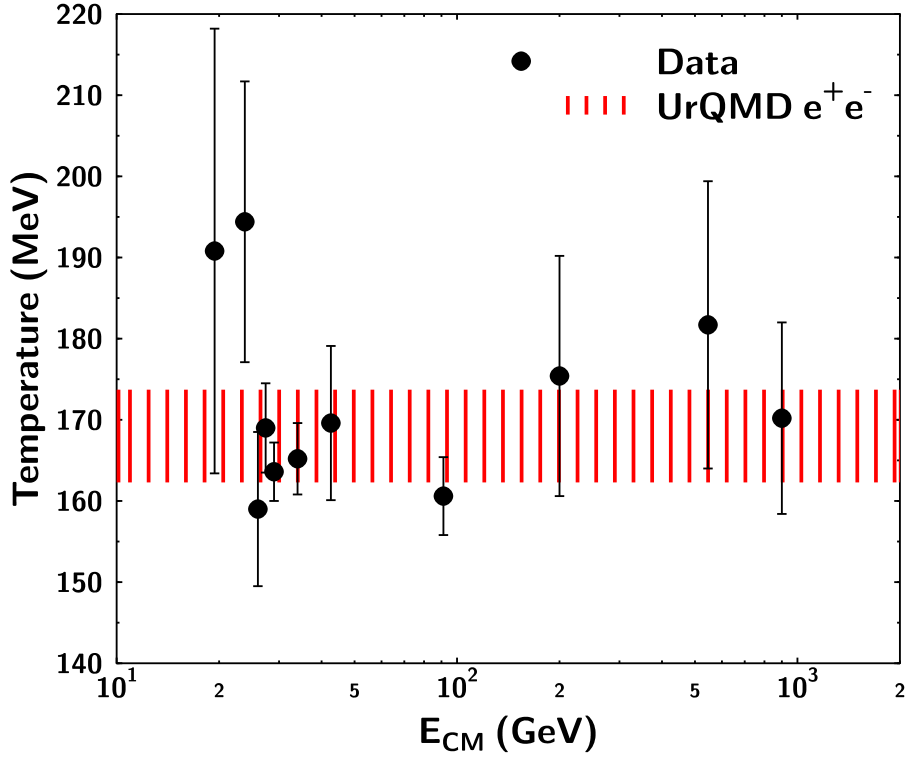,width=15cm}}
\caption{Pion 'temperatures' (inv. slope parameter of the $p_t$
distribution) extracted from $e^+e^-$ annihilations in the UrQMD model
for different energies are shown.
They are compared to freeze-out 'temperatures' extracted from a statistical
model fit to particle yields \protect\cite{becattini} in $pp$, $\overline p p$
and $e^+e^-$ reactions.
}
\label{fig28}
\end{figure}

The total energy and the masses of the incoming hadrons serve as
input for calculating the angular distribution.
It is worth to stress again that Eq.(\ref{angeq}) is used to calculate
only the
angular distributions for all elementary elastic two-body processes but
not the corresponding total cross sections.
The inverse slopes ('temperatures') as calculated in the UrQMD from
the transverse momentum spectra of pions - by fitting the $1/m_t dN/dm_t$
distribution with an exponential - compare well with values extracted from
thermal fits to data\footnote{Temperatures have been
extracted from a statistical model fit to the particle yields in $pp$,
$\overline p p$ and $e^+e^-$.} \cite{becattini} (Fig.~\ref{fig28}).

\section{Particle Yields, Longitudinal and Transverse Spectra}
\label{sec6}

The UrQMD model reproduces nicely the total, elastic and inelastic
cross-sections of numerous hadronic reactions. The model also predicts
the particle multiplicities (i.e. the inclusive
cross-sections) as well as the (Lorentz-invariant) cross-sections, which
may come in the form of $x_F$-, rapidity-, or transverse momentum
distributions. The abundances of the most important 
particle species produced in $pp$ collisions at 12 GeV/$c$ \cite{blobel}
are listed in Table~\ref{tab12gev} - the model predictions are in line
with the data within 15\%. The yields of various particle species produced in
$pp$ collisions at 205 GeV/$c$ ($\sqrt{s} \approx 19.7$ GeV)
\cite{kafka} are listed in Table~\ref{tab6}. It is easy to see that
the model predictions lie generally well within the 10\%
of the data except for the strange baryons. Table~\ref{tab7} presents
the results on particle production in $pp$ interactions at
$\sqrt{s} = 27$ GeV \cite{27gevdata}. Again, the agreement between the
experimental and theoretical results is good. As discussed above,
the production of $\Lambda$($\bar \Lambda$)'s is overestimated by a
factor of 2-3 due to the neglect of finite size effects in the strings.

\begin{figure}
\centerline{\psfig{figure=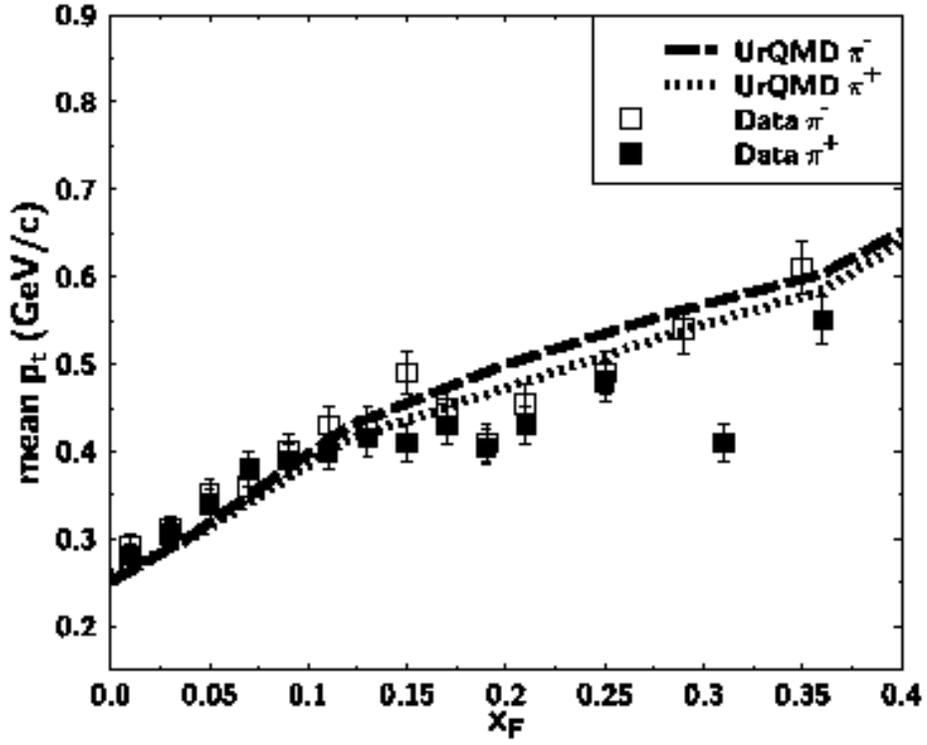,width=15cm}}
\caption{Mean transverse momentum of $\pi^+$ (dotted line) and
$\pi^-$ (dashed line) in $pp$ collisions at 205 GeV/$c$ as a function
of $x_F$. Data ($\pi^+$'s, open squares, and $\pi^-$'s, full squares)
are taken from \protect\cite{kafka}.
}
\label{fig29}
\end{figure}

\begin{figure}
\centerline{\psfig{figure=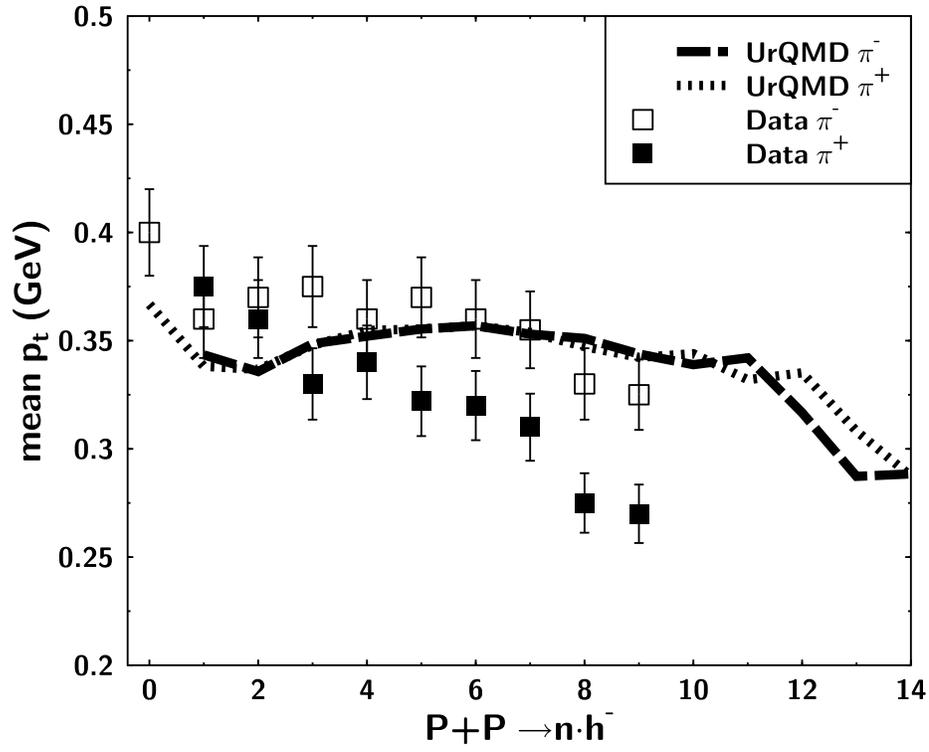,width=15cm}}
\caption{The mean
transverse momentum of $\pi^{\pm}$'s as a function of number of
negatively charged hadrons in the reaction p(205 GeV)+p. Note the
suppressed zero. Data are taken from \protect\cite{kafka}.
}
\label{fig30}
\end{figure}

The rapidity spectra of pions and $x_F$-distributions of baryons,
as obtained from the UrQMD model for $pp$ interactions at 205 GeV/$c$
(Fig.~\ref{fig22} and \ref{fig23}, respectively) have been discussed
already in the previous section.
Correlations between the transverse and longitudinal momenta
of charged pions, produced in the same reaction, can be studied \cite{kafka}.
The transverse
momenta of both positively and negatively charged pions have
evident minima at $x_F = 0$ as shown in Fig.~\ref{fig29} (the
''sea-gull'' effect). Then the transverse momentum increase nearly
linearly with the rising longitudinal momenta. The agreement
with the experimental data is fair. The single event
correlation between the transverse momenta of the produced pions and the
multiplicity of the negatively charged hadrons in the event is presented
in Fig.~\ref{fig30}. The measured distribution is reproduced
nicely for $\pi^+$'s , while for the $\pi^-$'s, 20\% deviations of the
calculated
spectrum from the data are evident at high ($n_{h^-} \geq 5$) event
multiplicities.

\begin{figure}
\centerline{\psfig{figure=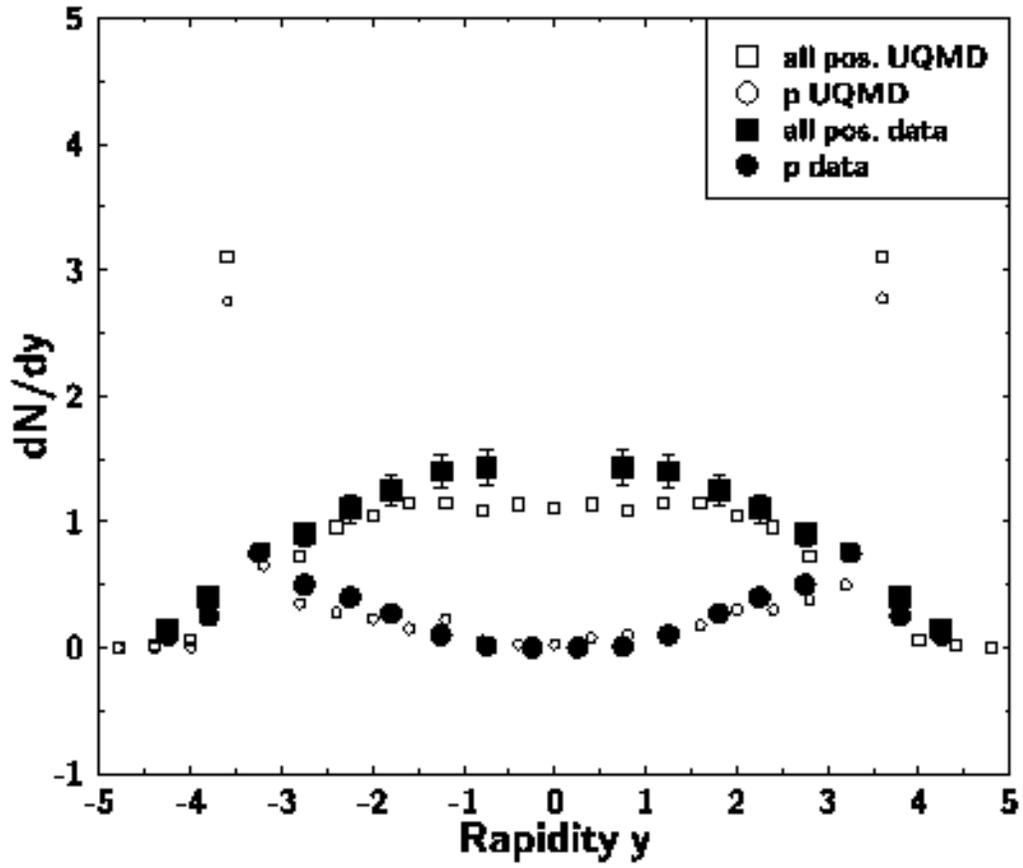,width=15cm}}
\caption{Rapidity distribution of protons and
positively charged particles for the reaction He+He 
at $\sqrt{s}=31$~AGeV compared to data \protect\cite{otterlund}.
}
\label{hehe}
\end{figure}

To probe the ability of the UrQMD model to describe hadron-hadron
collision even at the today highest available bombarding energies for  
nuclei, we compare the calculated He+He collision at ISR with
data \cite{otterlund} as shown in Fig.~\ref{hehe}.

It is not surprising, that the light helium system is transparent. 
A baryon free area of 3 units in rapidity is produced. 
The UrQMD model prediction describes the data fairly well. 
The calculated particle yields can be increased by 15\% if 
one includes multiple jet creation into the model description \cite{tom}. 

The UrQMD model reproduces the cross-sections and spectra
of particles in hadronic collisions fairly well. Since hadronic
interactions build up the basic input to simulate the hadron-nucleus
and nucleus-nucleus interactions in the model, it is interesting to
examine the applicability of the UrQMD model to these reactions. The full
comparison with the experimental $h A$ and $A A$ data is an
ongoing program, which is not completed yet
\cite{ongoing}.

\begin{figure}
\centerline{\psfig{figure=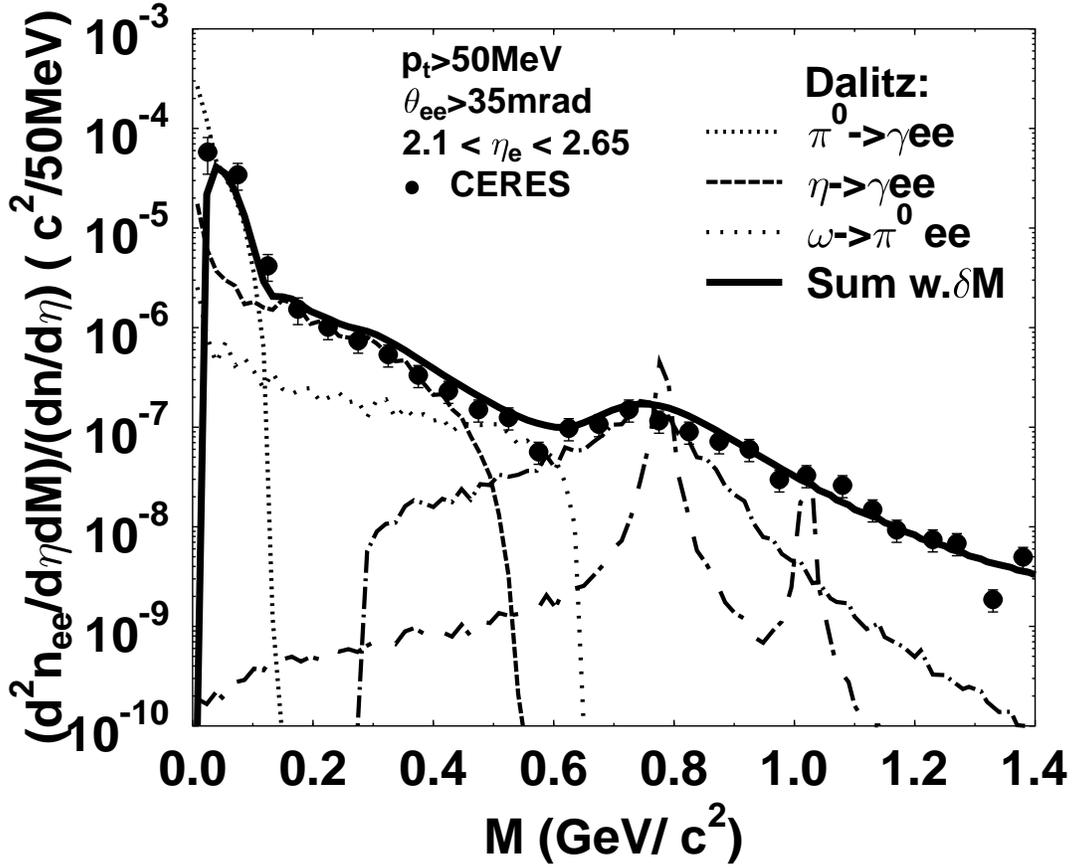,width=15cm}}
\caption{
Dilepton mass spectrum for $p$+Be at 450~GeV/$c$.
The calculation includes Dalitz decays and conversion
of vector mesons.
Only the curve labeled  sum of all contributions (solid curve) is folded with the
mass resolution of the CERES (full circles) experiment \protect\cite{specht}.
}
\label{fig31}
\end{figure}

The ability of the model to reproduce e.g.
dilepton yields in $pp$ collisions, which is of
interest in high energy physics is shown here as an example.
Fig.~\ref{fig31} shows the UrQMD calculations of the dilepton spectrum for
$p$+Be (which serves as a substitute for pp reactions)
at 450 GeV/$c$. Dilepton sources considered here are
Dalitz decays ($\pi^0$, $\eta$ and $\omega$) and direct vector meson decays
($\rho$, $\omega$ and $\phi$). Dalitz decays of heavier meson and
baryon resonances are included explicitly via their emission of
$\rho$ mesons (assuming vector meson dominance). To avoid
double counting, the $\rho$ mesons from $\eta$'s, and $\omega$'s are
excluded from the $\rho$ contribution. Pion annihilation is included
dynamically into the contribution of decaying $\rho$ mesons via the
channel $\pi^+\pi^- \rightarrow \rho \rightarrow e^+e^-$. The
calculation of dilepton yields without modifications of the $\rho$
mass pole is compatible with the CERES data \cite{specht}.

\section{Summary}

The basic hadronic interaction processes incorporated
into the UrQMD approach are given. The implemented
cross-sections of various hadron-hadron collisions, as well as their
extrapolations into the high energy region are presented. The model
treatment of
excitation and decay of intermediate objects like resonances and
strings is described. The importance of the finite size
effects for the process of string fragmentation is discussed.
The UrQMD model is
a microscopic transport model which allows to study optionally color
fluctuations, i.e. color coherence phenomena, as Color Transparency and
Color Opacity, as well as the expansion of small wave packets from the
point of the production. The model predictions are compared with the
available experimental data on particle yields in hadronic
interactions for a wide range of c.m. energies. The UrQMD model 
treats the elementary processes reasonably well.
More accurate data on proton-proton, proton-neutron, as well as
meson-baryon collisions are needed to improve the  extrapolation
to nucleus-nucleus interactions at high energies, where already in
the present model - about $10^4$ elementary
hadron-hadron reactions are possible.

\section*{Acknowledgements}

This work was supported by the Graduiertenkolleg f{\"u}r Theoretische
und Experimentelle Schwerionenphysik, Frankfurt$-$Giessen, the
Bundesministerium f{\"u}r Bildung und Forschung, the Gesellschaft
f{\"u}r Schwerionenforschung, Darmstadt, Deutsche
Forschungsgemeinschaft, the Alexander von Humboldt$-$Stiftung, the
Josef Buchmann-Stiftung, DAAD and in part by
DOE grant DE-FG02-96ER40945.

\widetext

\newpage

\newpage

\newpage

\newpage

\newpage

\newpage

\newpage

\newpage

\newpage

\newpage

\newpage

\newpage

\newpage

\newpage

\newpage

\newpage

\newpage

\newpage

\newpage

\newpage

\newpage

\newpage

\newpage

\newpage

\newpage

\newpage

\newpage

\newpage

\newpage


\newpage

\newpage

\newpage

\newpage

\newpage

\newpage

\newpage

\clearpage
\newpage
\mediumtext

\begin{table}

\caption{Baryons and baryon resonances implemented in the UrQMD model.
All baryons up to 2.25 GeV/$c^2$ as well as their antiparticles are
included.
}
\begin{tabular}{cccccc}
$N$(Nucleon)&$\Delta$(Delta)&$\Lambda$(Lambda)&$\Sigma$(Sigma)&
$\Xi$(Xi)&$\Omega$(Omega)\\
\tableline
$N_{938} $&$\Delta_{1232}$&$\Lambda_{1116}$&$\Sigma_{1192}$
&$\Xi_{1317}$&$\Omega_{1672}$\\
$N_{1440}$&$\Delta_{1600}$&$\Lambda_{1405}$&$\Sigma_{1385}$&$\Xi_{1530}$&\\
$N_{1520}$&$\Delta_{1620}$&$\Lambda_{1520}$&$\Sigma_{1660}$&$\Xi_{1690}$&\\
$N_{1535}$&$\Delta_{1700}$&$\Lambda_{1600}$&$\Sigma_{1670}$&$\Xi_{1820}$&\\
$N_{1650}$&$\Delta_{1900}$&$\Lambda_{1670}$&$\Sigma_{1775}$&$\Xi_{1950}$&\\
$N_{1675}$&$\Delta_{1905}$&$\Lambda_{1690}$&$\Sigma_{1790}$&$$&\\
$N_{1680}$&$\Delta_{1910}$&$\Lambda_{1800}$&$\Sigma_{1915}$&$$&\\
$N_{1700}$&$\Delta_{1920}$&$\Lambda_{1810}$&$\Sigma_{1940}$&$$&\\
$N_{1710}$&$\Delta_{1930}$&$\Lambda_{1820}$&$\Sigma_{2030}$&&\\
$N_{1720}$&$\Delta_{1950}$&$\Lambda_{1830}$&$$&&\\
$N_{1900}$&&$\Lambda_{2100}$&$$&&\\
$N_{1990}$&&$\Lambda_{2110}$&$$&& \\
$N_{2080}$ &&&&&\\
$N_{2190}$ &&&&&\\
$N_{2200}$ &&&&&\\
$N_{2250}$ &&&&&\\
\end{tabular}
\label{tab1}
\end{table}

\begin{table}
\caption{Parameters of the CERN-HERA fit \protect\cite{pdg96} 
used in UrQMD for the total
and elastic cross-section above the resonance region ($p_{lab}>2$~GeV/$c$).
The cross sections are parametrized as: 
$\sigma_{\rm tot,el}(p)\,=\,{\rm A} + {\rm B}\, p^n +
{\rm C}\,{\rm ln}^2(p) + {\rm D}\,{\rm ln}(p)$, 
with the laboratory momentum $p$ in GeV/$c$ and the cross-section
$\sigma$ in mb.}
\begin{tabular}{lccccc}
$\sigma$                  & A    & B    & C      & D     & n \\\tableline
$pp$ (total)              &$ 48.0 $&$ 0.   $&$ 0.522 $ &$ -4.51$ &$ 0.   $       \\
$pp$ (elastic)            &$ 11.9 $&$ 26.9 $&$ 0.169 $ &$ -1.85$ &$ -1.21$ \\
$pn$ (total)              &$ 47.3 $&$ 0.   $&$ 0.513 $ &$ -4.27$ &$ 0.   $    \\
$\overline p p$ (total)   &$ 38.4 $&$ 77.6 $&$ 0.26  $ &$ -1.2 $ &$ -0.64$ \\
$\overline p p$ (elastic) &$ 10.2 $&$ 52.7 $&$ 0.125 $ &$ -1.28$ &$ -1.16$ \\
$\gamma p$ (total)        &$ 0.147$&$ 0.   $&$ 0.0022$ &$ -.017$ &$ 0.   $    \\
$\pi^+ p$ (total)         &$ 16.4 $&$ 19.3 $&$ 0.19  $ &$ 0.   $ &$ -0.42$      \\
$\pi^+ p$ (elastic)       &$ 0.   $&$ 11.4 $&$ 0.079 $ &$ 0.   $ &$ -0.4 $        \\
$\pi^- p$ (total)         &$ 33.0 $&$ 14.0 $&$ 0.456 $ &$ -4.03$ &$ -1.36$ \\
$\pi^- p$ (elastic)       &$ 1.76 $&$ 11.2 $&$ 0.043 $ &$ 0.   $ &$ -0.64$   \\
$K^+ p$ (total)           &$ 18.1 $&$ 0.   $&$ 0.26  $ &$ -1.  $ &$ 0.   $        \\
$K^+ p$ (elastic)         &$ 5.0  $&$ 8.1  $&$ 0.16  $ &$ -1.3 $ &$ -1.8 $     \\
$K^+ n$ (total)           &$ 18.7 $&$ 0.   $&$ 0.21  $ &$ -0.89$ &$ 0.   $       \\
$K^- p$ (total)           &$ 32.1 $&$ 0.   $&$ 0.66  $ &$ -5.6 $ &$ 0.   $       \\
$K^- p$ (elastic)         &$ 7.3  $&$ 0.   $&$ 0.29  $ &$ -2.4 $ &$ 0.   $        \\
$K^- n$ (total)           &$ 25.2 $&$ 0.   $&$ 0.38  $ &$ -2.9 $ &$ 0.   $       \\
\end{tabular}
\label{tab2}
\end{table}

\begin{table}
\caption{
Baryon-baryon cross-sections  in [mb] from the Additive Quark Model.
NN scattering is explicitly treated, i.e. $\sqrt{s}$-dependent, etc.}
\begin{tabular}{cccccc}
B$_1$\ B$_2$ &  N  &  $\Lambda$  &  $\Xi$  &  $\Omega$  \\ \tableline
 $N$         &40.0 & 34.7        &  29.3   & 24.0       \\
 $\Lambda$   &34.7 & 30.0        &  25.4   & 20.8       \\
 $\Xi$       &29.3 & 25.4        & 21.5    & 17.6       \\
 $\Omega$    &24.0 & 20.8        & 17.6    & 14.4       \\
\end{tabular}
\label{tab3}
\end{table}

\begin{table}
\caption{
Meson-baryon cross-sections in [mb] from the Additive Quark Model.
MB scattering in the resonance region ($\sqrt s <1.7$~GeV) is
explicitly treated.}
\begin{tabular}{ccccc}
M$_1$\ B$_2$ &  N   &  $\Lambda$  &  $\Xi$  &  $\Omega$  \\ \hline
 $\pi$       & 26.6 & 23.1        & 19.6    & 16.0       \\
 $K$         & 21.3 & 18.5        & 15.6    & 12.8       \\
 $\Phi$      & 16.0 & 13.9        & 11.7    & 9.6        \\
\end{tabular}
\label{tab4}
\end{table}

\begin{table}
\caption{
Meson-Meson cross-sections  in [mb] from the Additive Quark Model.
MM scattering in the resonance region ($\sqrt s <1.7$~GeV) is
explicitly treated.}
\begin{tabular}{cccc}
M$_1$\ M$_2$ &  $\pi$  &  $K$  &  $\Phi$  \\ \hline
 $\pi$       & 17.8    & 14.2  & 10.7     \\
 $K$         & 14.2    & 11.4  & 8.5      \\
 $\Phi$      & 10.7    & 8.5   & 6.4      \\
\end{tabular}
\label{tab5}
\end{table}

\begin{table}
\caption{Mixing angles of meson multiplets according to the flavor SU$(3)$ quark
model: these parameters assign the pure $u\overline u$, $d\overline d$, 
$s\overline s$,to the physical particles according to the SU$(3)$ quark model.
The flavor mixing angles are chosen according to quadratic Gell-Mann-Okubo mass
formula \protect\cite{gellm}. For the scalar mesons this formula is not
applicable, here an ideal mixing angle (${\rm tan}(\theta)=1/\sqrt{2}$)
is assumed.
}
\begin{tabular}{lc}
Multiplet              &  degree      \\ \hline
 scalar                & 35   \\
 pseudoscalar          & -10  \\
 vector                & 39   \\
 pseudovector          & 51   \\
 tensor                & 28   \\
\end{tabular}
\label{mixang}
\end{table}

\begin{table}
\caption{Particle multiplicities from the UrQMD per inelastic
$pp$ event at 12 GeV/$c$. Data are taken from \protect\cite{blobel}.}
\begin{tabular}{lcc}
Particle  &   UrQMD       &      Exp. Data \\   
\tableline
$\pi^+$   &       1.22    &       1.44$\pm$0.02   \\
$\pi^-$   &       0.64    &       0.71$\pm$0.02   \\
$K^0_s$   &       0.019   &       0.019$\pm$0.001 \\
$p     $  &       1.38    &       1.27$\pm$0.02   \\
$\Lambda$ &       0.025   &       0.037$\pm$0.001 \\
\end{tabular}
\label{tab12gev}
\end{table}

\begin{table}
\caption{Particle multiplicities from the UrQMD per inelastic
$pp$ event at 205 GeV/$c$. Data are taken from \protect\cite{kafka}.}
\begin{tabular}{lcclcc}
Particle &Exp. Data  & UrQMD & Particle &Exp. Data  & UrQMD \\
\tableline
$\pi^-$    & 2.62$\pm$0.06 & 2.57 & $\pi^+$  &3.22$\pm$0.12 & 3.10 \\
$\pi^0$    & 3.34$\pm$0.24 & 3.11 & $K^+$    &0.28$\pm$0.06 & 0.26 \\
$K^-$      & 0.18$\pm$0.05 & 0.16 & $K^0$    &              & 0.24 \\
$\bar K^0$ &               & 0.16 & $K^0_S$  &0.17$\pm$0.01 & 0.20 \\
$\Lambda + \Sigma_0$       &0.096$\pm$0.01 & 0.16 &
$\bar \Lambda + \bar \Sigma^0$  &0.013$\pm$0.004& 0.037  \\
$p$        &1.34$\pm$0.15  & 1.32 & $\bar p$ &0.05$\pm$0.02 & 0.06 \\
\end{tabular}
\label{tab6}
\end{table}

\newpage
\begin{table}
\caption{Particle multiplicities from the UrQMD per inelastic $pp$
event at $\sqrt{s}=27\,$GeV. Data are taken from
\protect\cite{27gevdata}.}
\begin{tabular}{lcclcc}
Particle & Exp. Data & UrQMD & Particle & Exp. Data & UrQMD \\
\tableline
$\pi^+$  &4.10$\pm$ 0.26  & 3.79 & $\pi^0$  &3.87$\pm$ 0.28  & 3.72\\
$\pi^-$  &3.34$\pm$ 0.20  & 3.16 & $K^+$    &0.33$\pm$ 0.023 & 0.31\\
$K^-$    &0.22$\pm$ 0.015 & 0.22 & $K^0_S$  &0.23$\pm$ 0.015 & 0.26\\
$\eta$   &0.39$\pm$ 0.075 & 0.36 & $\rho_0$ &0.385$\pm$0.056 & 0.50\\
$\rho^+$ &0.552$\pm$0.129 & 0.52 & $\rho^-$ &0.355$\pm$0.091 & 0.41\\
$\omega$ &0.39$\pm$ 0.026 & 0.47 &$K^{\ast +}$&0.132$\pm$0.018&0.13\\
$K^{\ast -}$ &0.088$\pm$0.013 & 0.080 &
$K^{\ast 0}$ &0.119$\pm$0.023 & 0.123\\
$\bar K^{\ast 0}$ &0.09$\pm$ 0.017 & 0.081 &
$\phi$ &0.019$\pm$0.002 & 0.009\\
$f_2(1270)$ &0.092$\pm$0.013 & 0.119 &
$p$   &1.20$ \pm$0.119 & 1.32\\
$\bar p$ &0.063$\pm$0.003 & 0.088 &
$\Lambda + \Sigma^0$ &0.125$\pm$0.016 & 0.19\\
$\Lambda$      &             & 0.15 &
$\Sigma^0$   &             & 0.041\\
$\bar \Lambda + \bar \Sigma^0$ &0.020$\pm$0.005 & 0.047 &
$\bar \Lambda$       &             & 0.038\\

$\bar \Sigma^0$ &             & 0.009 &
$\Sigma^+$   &0.048$\pm$0.019 & 0.050\\
$\Sigma^0$   &           & 0.041 &
$\Xi^-$      &            & 0.0041\\
$\Xi^+$      &            & 0.0053 &
$\Sigma^-$   &0.013$\pm$0.009 & 0.015\\
$\Delta^{++}$  &0.218$\pm$0.016 & 0.235 &
$\Delta^0$   &0.141$\pm$0.019 & 0.197\\
$\bar \Delta^{++}$ &0.013$\pm$0.005 & 0.016 &
$\bar \Delta^0$  &0.034$\pm$0.009 & 0.026\\
$\Sigma^{\ast +}$  &0.020$\pm$0.004 & 0.040 &
$\Sigma^{\ast 0}$  &            & 0.071\\
$\Sigma^{\ast -}$  &0.010$\pm$0.003 & 0.009 &  &  & \\
\end{tabular}
\label{tab7}
\end{table}

\end{document}